%% file: main.tex
\documentclass{aastex631}

\usepackage[T5,T1]{fontenc}
\usepackage[all]{hypcap}
\usepackage[section]{placeins}

\graphicspath{{./}{figures/}}

\begin{document}

\title{ Search for neutrino emission from hard X-ray AGN with IceCube}


\affiliation{III. Physikalisches Institut, RWTH Aachen University, D-52056 Aachen, Germany}
\affiliation{Department of Physics, University of Adelaide, Adelaide, 5005, Australia}
\affiliation{Dept. of Physics and Astronomy, University of Alaska Anchorage, 3211 Providence Dr., Anchorage, AK 99508, USA}
\affiliation{Dept. of Physics, University of Texas at Arlington, 502 Yates St., Science Hall Rm 108, Box 19059, Arlington, TX 76019, USA}
\affiliation{CTSPS, Clark-Atlanta University, Atlanta, GA 30314, USA}
\affiliation{School of Physics and Center for Relativistic Astrophysics, Georgia Institute of Technology, Atlanta, GA 30332, USA}
\affiliation{Dept. of Physics, Southern University, Baton Rouge, LA 70813, USA}
\affiliation{Dept. of Physics, University of California, Berkeley, CA 94720, USA}
\affiliation{Lawrence Berkeley National Laboratory, Berkeley, CA 94720, USA}
\affiliation{Institut f{\"u}r Physik, Humboldt-Universit{\"a}t zu Berlin, D-12489 Berlin, Germany}
\affiliation{Fakult{\"a}t f{\"u}r Physik {\&} Astronomie, Ruhr-Universit{\"a}t Bochum, D-44780 Bochum, Germany}
\affiliation{Universit{\'e} Libre de Bruxelles, Science Faculty CP230, B-1050 Brussels, Belgium}
\affiliation{Vrije Universiteit Brussel (VUB), Dienst ELEM, B-1050 Brussels, Belgium}
\affiliation{Department of Physics and Laboratory for Particle Physics and Cosmology, Harvard University, Cambridge, MA 02138, USA}
\affiliation{Dept. of Physics, Massachusetts Institute of Technology, Cambridge, MA 02139, USA}
\affiliation{Dept. of Physics and The International Center for Hadron Astrophysics, Chiba University, Chiba 263-8522, Japan}
\affiliation{Department of Physics, Loyola University Chicago, Chicago, IL 60660, USA}
\affiliation{Dept. of Physics and Astronomy, University of Canterbury, Private Bag 4800, Christchurch, New Zealand}
\affiliation{Dept. of Physics, University of Maryland, College Park, MD 20742, USA}
\affiliation{Dept. of Astronomy, Ohio State University, Columbus, OH 43210, USA}
\affiliation{Dept. of Physics and Center for Cosmology and Astro-Particle Physics, Ohio State University, Columbus, OH 43210, USA}
\affiliation{Niels Bohr Institute, University of Copenhagen, DK-2100 Copenhagen, Denmark}
\affiliation{Dept. of Physics, TU Dortmund University, D-44221 Dortmund, Germany}
\affiliation{Dept. of Physics and Astronomy, Michigan State University, East Lansing, MI 48824, USA}
\affiliation{Dept. of Physics, University of Alberta, Edmonton, Alberta, T6G 2E1, Canada}
\affiliation{Erlangen Centre for Astroparticle Physics, Friedrich-Alexander-Universit{\"a}t Erlangen-N{\"u}rnberg, D-91058 Erlangen, Germany}
\affiliation{Physik-department, Technische Universit{\"a}t M{\"u}nchen, D-85748 Garching, Germany}
\affiliation{D{\'e}partement de physique nucl{\'e}aire et corpusculaire, Universit{\'e} de Gen{\`e}ve, CH-1211 Gen{\`e}ve, Switzerland}
\affiliation{Dept. of Physics and Astronomy, University of Gent, B-9000 Gent, Belgium}
\affiliation{Dept. of Physics and Astronomy, University of California, Irvine, CA 92697, USA}
\affiliation{Karlsruhe Institute of Technology, Institute for Astroparticle Physics, D-76021 Karlsruhe, Germany}
\affiliation{Karlsruhe Institute of Technology, Institute of Experimental Particle Physics, D-76021 Karlsruhe, Germany}
\affiliation{Dept. of Physics, Engineering Physics, and Astronomy, Queen's University, Kingston, ON K7L 3N6, Canada}
\affiliation{Department of Physics {\&} Astronomy, University of Nevada, Las Vegas, NV 89154, USA}
\affiliation{Nevada Center for Astrophysics, University of Nevada, Las Vegas, NV 89154, USA}
\affiliation{Dept. of Physics and Astronomy, University of Kansas, Lawrence, KS 66045, USA}
\affiliation{Centre for Cosmology, Particle Physics and Phenomenology - CP3, Universit{\'e} catholique de Louvain, Louvain-la-Neuve, Belgium}
\affiliation{Department of Physics, Mercer University, Macon, GA 31207-0001, USA}
\affiliation{Dept. of Astronomy, University of Wisconsin{\textemdash}Madison, Madison, WI 53706, USA}
\affiliation{Dept. of Physics and Wisconsin IceCube Particle Astrophysics Center, University of Wisconsin{\textemdash}Madison, Madison, WI 53706, USA}
\affiliation{Institute of Physics, University of Mainz, Staudinger Weg 7, D-55099 Mainz, Germany}
\affiliation{Department of Physics, Marquette University, Milwaukee, WI 53201, USA}
\affiliation{Institut f{\"u}r Kernphysik, Westf{\"a}lische Wilhelms-Universit{\"a}t M{\"u}nster, D-48149 M{\"u}nster, Germany}
\affiliation{Bartol Research Institute and Dept. of Physics and Astronomy, University of Delaware, Newark, DE 19716, USA}
\affiliation{Dept. of Physics, Yale University, New Haven, CT 06520, USA}
\affiliation{Columbia Astrophysics and Nevis Laboratories, Columbia University, New York, NY 10027, USA}
\affiliation{Dept. of Physics, University of Oxford, Parks Road, Oxford OX1 3PU, United Kingdom}
\affiliation{Dipartimento di Fisica e Astronomia Galileo Galilei, Universit{\`a} Degli Studi di Padova, I-35122 Padova PD, Italy}
\affiliation{Dept. of Physics, Drexel University, 3141 Chestnut Street, Philadelphia, PA 19104, USA}
\affiliation{Physics Department, South Dakota School of Mines and Technology, Rapid City, SD 57701, USA}
\affiliation{Dept. of Physics, University of Wisconsin, River Falls, WI 54022, USA}
\affiliation{Dept. of Physics and Astronomy, University of Rochester, Rochester, NY 14627, USA}
\affiliation{Department of Physics and Astronomy, University of Utah, Salt Lake City, UT 84112, USA}
\affiliation{Dept. of Physics, Chung-Ang University, Seoul 06974, Republic of Korea}
\affiliation{Oskar Klein Centre and Dept. of Physics, Stockholm University, SE-10691 Stockholm, Sweden}
\affiliation{Dept. of Physics and Astronomy, Stony Brook University, Stony Brook, NY 11794-3800, USA}
\affiliation{Dept. of Physics, Sungkyunkwan University, Suwon 16419, Republic of Korea}
\affiliation{Institute of Basic Science, Sungkyunkwan University, Suwon 16419, Republic of Korea}
\affiliation{Institute of Physics, Academia Sinica, Taipei, 11529, Taiwan}
\affiliation{Dept. of Physics and Astronomy, University of Alabama, Tuscaloosa, AL 35487, USA}
\affiliation{Dept. of Astronomy and Astrophysics, Pennsylvania State University, University Park, PA 16802, USA}
\affiliation{Dept. of Physics, Pennsylvania State University, University Park, PA 16802, USA}
\affiliation{Dept. of Physics and Astronomy, Uppsala University, Box 516, SE-75120 Uppsala, Sweden}
\affiliation{Dept. of Physics, University of Wuppertal, D-42119 Wuppertal, Germany}
\affiliation{Deutsches Elektronen-Synchrotron DESY, Platanenallee 6, D-15738 Zeuthen, Germany}

\author[0000-0001-6141-4205]{R. Abbasi}
\affiliation{Department of Physics, Loyola University Chicago, Chicago, IL 60660, USA}

\author[0000-0001-8952-588X]{M. Ackermann}
\affiliation{Deutsches Elektronen-Synchrotron DESY, Platanenallee 6, D-15738 Zeuthen, Germany}

\author{J. Adams}
\affiliation{Dept. of Physics and Astronomy, University of Canterbury, Private Bag 4800, Christchurch, New Zealand}

\author[0000-0002-9714-8866]{S. K. Agarwalla}
\altaffiliation{also at Institute of Physics, Sachivalaya Marg, Sainik School Post, Bhubaneswar 751005, India}
\affiliation{Dept. of Physics and Wisconsin IceCube Particle Astrophysics Center, University of Wisconsin{\textemdash}Madison, Madison, WI 53706, USA}

\author[0000-0003-2252-9514]{J. A. Aguilar}
\affiliation{Universit{\'e} Libre de Bruxelles, Science Faculty CP230, B-1050 Brussels, Belgium}

\author[0000-0003-0709-5631]{M. Ahlers}
\affiliation{Niels Bohr Institute, University of Copenhagen, DK-2100 Copenhagen, Denmark}

\author[0000-0002-9534-9189]{J.M. Alameddine}
\affiliation{Dept. of Physics, TU Dortmund University, D-44221 Dortmund, Germany}

\author{N. M. Amin}
\affiliation{Bartol Research Institute and Dept. of Physics and Astronomy, University of Delaware, Newark, DE 19716, USA}

\author[0000-0001-9394-0007]{K. Andeen}
\affiliation{Department of Physics, Marquette University, Milwaukee, WI 53201, USA}

\author[0000-0003-4186-4182]{C. Arg{\"u}elles}
\affiliation{Department of Physics and Laboratory for Particle Physics and Cosmology, Harvard University, Cambridge, MA 02138, USA}

% ...


\collaboration{426}{IceCube Collaboration}



\begin{abstract}

Active Galactic Nuclei (AGN) are promising candidate sources of
high-energy astrophysical neutrinos since they provide environments rich
in matter and photon targets where cosmic ray interactions may lead to
the production of gamma rays and neutrinos. We searched for high-energy neutrino emission from AGN using the \textit{Swift}-BAT Spectroscopic Survey (BASS) catalog of hard X-ray sources and 12~years of IceCube muon track data. First, upon performing a stacked search, no significant emission was found. Second, we searched for neutrinos from a list of 43 candidate sources and found an excess from the direction of two sources, Seyfert galaxies NGC~1068 and NGC~4151. We observed NGC~1068 at flux $\phi_{\nu_{\mu}+\bar{\nu}_{\mu}}$ =  $4.02_{-1.52}^{+1.58} \times 10^{-11}$~TeV$^{-1}$~cm$^{-2}$~s$^{-1}$ normalized at 1~TeV, with power-law spectral index, $\gamma$ = 3.10$^{+0.26}_{-0.22}$, consistent with previous IceCube results. The observation of a neutrino excess from the direction of NGC~4151 is at a post-trial significance of 2.9$\sigma$. If interpreted as an astrophysical signal, the excess observed from NGC~4151 corresponds to a flux $\phi_{\nu_{\mu}+\bar{\nu}_{\mu}}$ = $1.51_{-0.81}^{+0.99} \times 10^{-11}$~TeV$^{-1}$~cm$^{-2}$~s$^{-1}$ normalized at 1~TeV and $\gamma$ = 2.83$^{+0.35}_{-0.28}$.

\end{abstract}


\section{Introduction}

In 2013, IceCube reported the detection of high-energy neutrinos of astrophysical origin with a sky distribution consistent with isotropy~\citep{IceCube:2013low}. This steady flux of astrophysical neutrinos observed in the energy range  TeV-PeV has so far been characterized by a single power-law model \citep{ PhysRevLett.125.121104, 2020arXiv201103561A, PhysRevD.104.022002, Abbasi_2022}. To find the origin of the neutrino emission, IceCube has searched for neutrino excesses from various astronomical source classes, for e.g., supernovae~\citep{IceCube:2023esf}, pulsar wind nebulae~\citep{Aartsen_2020}, ultra-luminous infrared galaxies~\citep{Correa_2021}, X-ray binaries~\citep{IceCube:2022jpz}, galaxy clusters~\citep{Abbasi_2022_gc}, GRBs~\citep{IceCube:2022rlk} and others. Evidence of neutrinos from the galactic plane~\citep{doi:10.1126/science.adc9818}, at a significance level of 4.5$\sigma$, indicates a contribution of $\sim$ 6\% to 13\% to the all-sky astrophysical flux at 30~TeV. So far, no significant evidence of a dominant source class has been identified.

Active galactic nuclei (AGN) are promising source candidates owing to their electromagnetic emission~\citep{10.1111/j.1365-2966.2012.21513.x} that spans several orders of magnitude in luminosity and likely ability to accelerate charged particles to ultra-high energies, $E\sim$10$^{20}$~eV \citep{Mbarek_2019}. AGN  are the compact innermost regions of galaxies, bright enough to outshine the host galaxy. They emit vast amounts of energy as matter spirals into a central supermassive black hole (SMBH). The emission of AGN spans the entire electromagnetic spectrum, from radio waves to gamma rays~\citep{1996ApJ...470..364E}. The AGN core is enveloped in a dusty torus and may be accompanied by relativistic jets or winds~\citep{2006MmSAI..77..598K}. The orientation of the jet with respect to the observer and the degree of obscuration by the dusty torus results in differences in the observed spectral features~\citep{1993ARA&A..31..473A, Urry_1995}. Blazars are the brightest AGN, with their jets pointing towards the Earth.

AGN have long been considered as production sites of astrophysical neutrinos~\citep[see, e.g.,][]{PhysRevLett.66.2697, MANNHEIM1995295, 1997ApJ...488..669H}. This hypothesis is supported by the detection of a high-energy neutrino in 2017 by IceCube in spatial and temporal coincidence with the blazar TXS 0506+056, in a flaring state \citep{doi:10.1126/science.aat2890}, and the evidence of neutrinos from the active galaxy NGC~1068 \citep{IceCube:2018dnn, doi:10.1126/science.abg3395}. Neutrino emission models suggest that high-energy neutrinos produced in AGN by the decay of charged pions are accompanied by a high-energy gamma-ray flux arising from the decay of neutral pions~\citep[see, e.g.,][]{Gao_2017, Keivani_2018, Murase:2022feu}. Previous studies have searched for neutrino emission from gamma-ray bright blazars  \citep[see, e.g.,][]{2017ApJ...835...45A,Huber:2019lrm} and concluded that this source class contributes only a small fraction towards the total flux of astrophysical neutrinos observed by IceCube. The lack of a simple association between gamma-ray AGN and neutrinos motivated new models \citep[see, e.g.,][]{Gao:2018mnu, Petropoulou_2020, Padovani:2023nei} where neutrinos escape the AGN environment but gamma rays interact with matter or photon fields surrounding the active core and cascade down to MeV gamma rays or hard X-rays. These predictions led to searches for neutrino emission from sources detected in lower energy bands which found background-compatible results \citep[see, e.g.,][]{IceCube:2022zbd, IceCube:2021pgw}.

In this study, we search for a possible correlation between high-energy neutrinos and AGN detected in hard X-rays. We use the \textit{Swift}-BAT AGN Spectroscopic Survey\footnote{\href{https://www.bass-survey.com/}{\textit{Swift}-BAT AGN Spectroscopic Survey}} \citep{2018AAS...23132001K} (BASS), which is the most complete all-sky catalog of hard X-ray AGN detected in the 14 - 195~keV range, and perform two different analyses as detailed in Section~\ref{subsec:analysis I} and Section~\ref{subsec:analysis II}.


\section{Dataset and Hard X-ray sources} \label{sec:dataset_n_catalog}

\subsection{The Neutrino Dataset} \label{sec:dataset}

The IceCube Neutrino Observatory is situated at the Amundsen-Scott South Pole Station encompassing a cubic kilometer of ice \citep{Aartsen_2017} that acts as the detection medium. Neutrino interactions produce secondary charged particles that move at relativistic speeds, emitting Cherenkov radiation. IceCube detects this radiation using 5160 digital optical modules (DOMs) deployed on 86 cables known as ``strings'' at depths between 1450~m and 2450~m  from the surface forming a three-dimensional array \citep{IceCube:2016zyt}. The events detected by IceCube are classified based on their optical signatures. The $\sim$km long track-like events are produced by charged-current (CC) interactions of muon neutrinos~\citep{Aartsen_2014}. The roughly spherical cascade events are produced by all other CC interactions and neutral current (NC) interactions~\citep{PhysRevLett.125.121104}. The background for the astrophysical neutrino flux is dominated by the atmospheric muons and neutrinos produced in cosmic-ray air showers~\citep{IceCube:2015wro}.

We used IceCube data recorded from 2008 to 2020 for the analyses discussed in this paper. The dataset has the same selection criteria as in~\citealp{PhysRevLett.124.051103} applied to an additional two years of data. The detector operated with partial configurations of 40, 59 and 79 strings from 2008 to 2011 and with the full 86-string configuration from 2011 onwards (see Table~\ref{tab:Dataset_info}). The sample covers the entire sky and is largely made of through-going muon tracks~\citep{PhysRevLett.124.051103}. The dataset was chosen due to the tracks having a better angular resolution compared to other event morphologies, with a median < 1$^{\circ}$ above 1 TeV ~\citep{Aartsen_2017, Abbasi_2021}. See~\citealp{PhysRevLett.124.051103} for more information about the selection criteria and the reconstruction methods for direction and energy.

\subsection{Hard X-ray Sources} \label{sec:source_catalog}
The BASS catalog~\citep{Ricci_2017, 2017ApJ...850...74K} used in this study contains AGN selected from sources in the 70-month \textit{ Swift}-BAT hard X-ray survey~\citep{Baumgartner_2013}. As the most complete hard X-ray AGN catalog, this provides an opportunity to perform a sensitive study of the potential contribution of hard X-ray AGN to the neutrino flux. Only six of the 838 sources do not have redshift estimates. Most are nearby, with a median of 0.04. We used the updated source class and redshift values from the BASS DR-2 catalog~\citep{Koss_2022, Koss_2022_cat}. 

Gas and dust surrounding AGN can act as targets for hadronic interactions. The amount of material, therefore, may impact neutrino production and is quantified by column density ($N_{\text{H}}$) of neutral hydrogen along the line of sight. Due to their high penetrating power, hard X-rays are well-suited to probe the obscured AGN core. Only two of the 838 sources do not have column density estimates. Since we have used column density, $N_{\text{H}}$, to classify the sources in the first analysis (see  Section~\ref{subsec:analysis I}), the two sources without the parameter measured were excluded.

\section{Analysis} \label{sec:analysis}

We perform two analyses: a stacked search and an individual source search, described in Section~\ref{subsec:analysis I} and Section~\ref{subsec:analysis II}, respectively. The analyses in this paper are time-integrated searches performed using an unbinned likelihood ratio method which tests how well a hypothesis representing a combination of the signal and the diffuse background describes the data as compared to a null hypothesis representing only the background~\citep{ACHTERBERG2006282, BRAUN2008299}. We assume a point source emits neutrinos following a power-law energy spectrum, with the neutrino flux per flavor given by, $\phi_{\nu_{\mu}+\bar{\nu}_{\mu}} (E_{\nu})= \phi_0 \cdot (E_{\nu}/E_0)^{-\gamma}$, where $E_0$ is the normalization energy, $\phi_0$ is the normalization flux, $E_{\nu}$ is the neutrino energy and spectral index $\gamma$. The likelihood function is defined as follows for a single point source~\citep{IceCube:2018ndw}:

\begin{equation}\label{eq:Likelihood}
\mathcal{L} (n_s, \gamma) = \prod_i^{N} \left(\frac{n_s}{N} \mathcal{S}_{i}(\vec{x_i},\sigma_i, E_i, \gamma) + \left(1-\frac{n_s}{N}\right) \mathcal{B}_{i}(\delta_i,E_i)\right) 
\end{equation}

where $n_s$ is the number of signal neutrinos. $\mathcal{S}_{i}$, the signal probability density functions (PDFs) have spatial and energy components where $\vec{x}_i$, $E_i$ and $\sigma_i$ are the reconstructed direction, energy and estimated directional uncertainty of each candidate neutrino event in the dataset, respectively. The spatial part of $\mathcal{B}_{i}$, the background PDF depends only on declination since the distribution of events in right ascension (RA) is uniform over long time scales for IceCube. The signal PDF is constructed using Monte Carlo (MC) simulated events and the background PDF using data randomized in RA. The number of signal events $n_s$ and the power-law index $\gamma$ are determined by maximizing the likelihood $\mathcal{L}$, which yields best-fit parameter values, $\hat{n}_s$ and $\hat{\gamma}$.

\subsection{The Stacked Search} \label{subsec:analysis I}

A stacking analysis evaluates the cumulative neutrino signal from a given population of sources. In this method, the signal from all sources is summed to increase the sensitivity and possibility of detection. The likelihood function used in this analysis is similar to Equation \ref{eq:Likelihood} but with a stacked signal PDF where each source term contains two types of weights:
\begin{itemize}
    \item \textbf{$\omega$:} a theoretical weight, which is a chosen observed physical property hypothesized to be proportional to the neutrino flux. In this analysis, we use the intrinsic X-ray flux (see Appendix \ref{sec:appendix_sources}) in the 14 - 195 keV range for each source as a weight. The values are provided in the BASS catalog.
    \item \textbf{$R(\delta_k,\gamma)$:} a detector weight, which is the sensitivity of the IceCube detector for a source at a declination $\delta $ with a spectral index $\gamma$.
\end{itemize}

The stacked signal PDF for $i$-th neutrino event, where $i \in \lbrace 1,2,...,N \rbrace$, $k \in \lbrace 1,2,...,M\rbrace$, $\vec{x}_{S_k}$ is the position of $k$-th source, and $M$ is the number of sources, is
\begin{equation}
    \mathcal{S}_{i,Stacked} =
      \frac{\sum_{k=1}^M \omega^k R^k(\delta_k,\gamma) \cdot
        \mathcal{S}_i^k(\vec{x}_i,\vec{x}_{S_k},\sigma_i,E_i, \gamma)}{\sum_{k=1}^M \omega^k R^k(\delta_k,\gamma)}. 
\end{equation}

For this analysis, we divide our catalog using two different criteria: AGN type and column density. We divide the catalog into blazar and non-blazar AGN according to the first criterion. Sub-dividing the catalog by the column density ($N_{\text{H}}$) results in three samples: unobscured AGN ($N_{\text{H}}$~<~10$^{22}$~cm$^{-2}$), obscured AGN (10$^{22}$ cm$^{-2}$~<~$N_{\text{H}}$~<~10$^{24}$~cm$^{-2}$) and Compton-thick (CT) AGN ($N_{\text{H}}$~>~10$^{24}$~cm$^{-2}$). Enough absorbing matter surrounds Compton-thick sources, resulting in a large optical depth for Compton scattering. We also test an unphysical hypothesis of equal neutrino emission from all AGN as an unbiased basis for comparison with other tests. In effect, we test a total of seven hypotheses. The first six use flux weights for each of the following AGN classes: (i)~all 836 AGN, (ii)~104 blazars, (iii)~732 non-blazars, (iv)~457 unobscured AGN, (v)~323 obscured AGN, (vi)~56 Compton-thick AGN, and the last (vii)~is using equal weights for all 836 AGN.

Any flux measurement (or flux upper limit) derived from the stacking analysis applies solely to the selected AGN of a given class that forms a subset of the entire source population in the universe.  Many AGN  fall below the instrument's detection threshold due to large distance, low intrinsic luminosity, or high obscuration. To consider all the undetected sources, we scale the fluxes by a catalog completeness factor.
In evaluating the contribution from the entire source population in the universe, we integrate over luminosity functions for each source class (blazars from~\citealp{BlazarCatComp} and non-blazar AGN from~\citealp{2014ApJ...786..104U}) and assume the neutrino flux is proportional to the intrinsic X-ray flux. See Appendix~\ref{sec:appendix_catalog_correction}, for the details of this method and the estimated catalog completeness factors for each AGN class.

\subsection{The Individual Source Search} \label{subsec:analysis II}

In this analysis, we study sources that are most likely to be individually detected by IceCube. We hypothesize that neutrino emission is proportional to the intrinsic hard X-ray flux in the 14 - 195~keV range. The strong declination dependence of the IceCube sensitivity to neutrino point sources implies that some sources have a higher likelihood of being detected because of their sky location.

To select the sources to be probed individually, we define a ``figure of merit'' (FOM) as the ratio of the intrinsic hard X-ray flux in the 14 - 195~keV range to the neutrino flux sensitivity at the source declination and rank all AGN in our catalog according to this value. Since a power-law energy spectrum with spectral index $\gamma = 2.5$ approximately describes the astrophysical diffuse flux, we considered the same for the neutrino flux from each source.  Starting from the source with the highest value of FOM, we select all the sources that have up to a factor of ten lower FOM than the top source (see Appendix~\ref{sec:appendix_point_source_selection}). The 43 AGN thus selected from the BASS catalog used to perform the second analysis are listed in Table~\ref{tab:catalog_search_upperlimits}. Most of these selected sources are in the Northern Hemisphere since IceCube has better sensitivity in the North as compared to the Southern sky. 

Using the method described in  Section~\ref{sec:analysis}, we find the best-fit values of $n_s$ and $\gamma$ for each source. NGC~1068 is present in the list used for this analysis. Since a previous IceCube study~\citep{doi:10.1126/science.abg3395} has identified evidence for neutrino emission at a significance of 4.2$\sigma$,  we do not report a post-trials significance for it.

\section{Results}\label{sec:results}
\textit{Stacking Analysis} -  We tested seven hypotheses: first, by using the entire AGN sample and five sub-samples with flux weights and second, by all AGN using equal weights. For each hypothesis, we computed pre-trials $p$-value ($p_{\text{local}}$). Since we test multiple hypotheses by performing several statistical tests and using samples having considerable overlaps, the $p_{\text{local}}$ for the most significant test is corrected using a trial factor (see Appendix~\ref{sec:appendix_trial_correc}). For the analysis performed using all 836 AGN with equal weights, the best-fit number of neutrinos is zero and $p_{\text{local}}$ is 1.0.  The hypothesis showing the highest statistical significance is obscured AGN  (10$^{22}$~cm$^{-2}$ < $N_{\text{H}}$ < 10$^{24}$~cm$^{-2}$) at a post-trials significance of 2.1$\sigma$. 

\begin{figure}[ht!]
\gridline{\fig{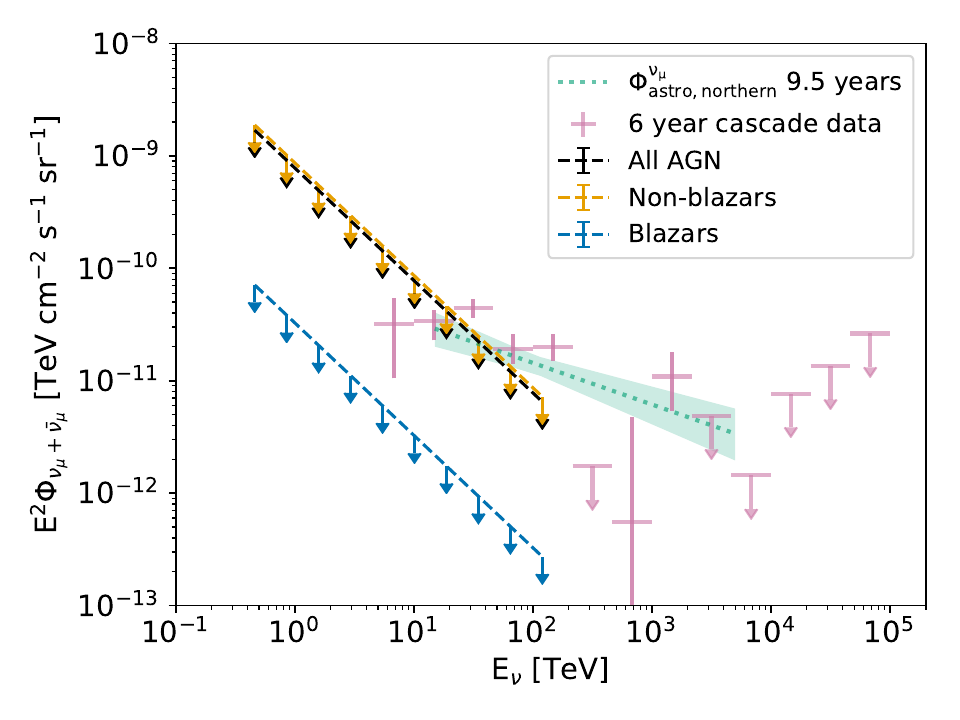}{0.42\linewidth}{}
          \fig{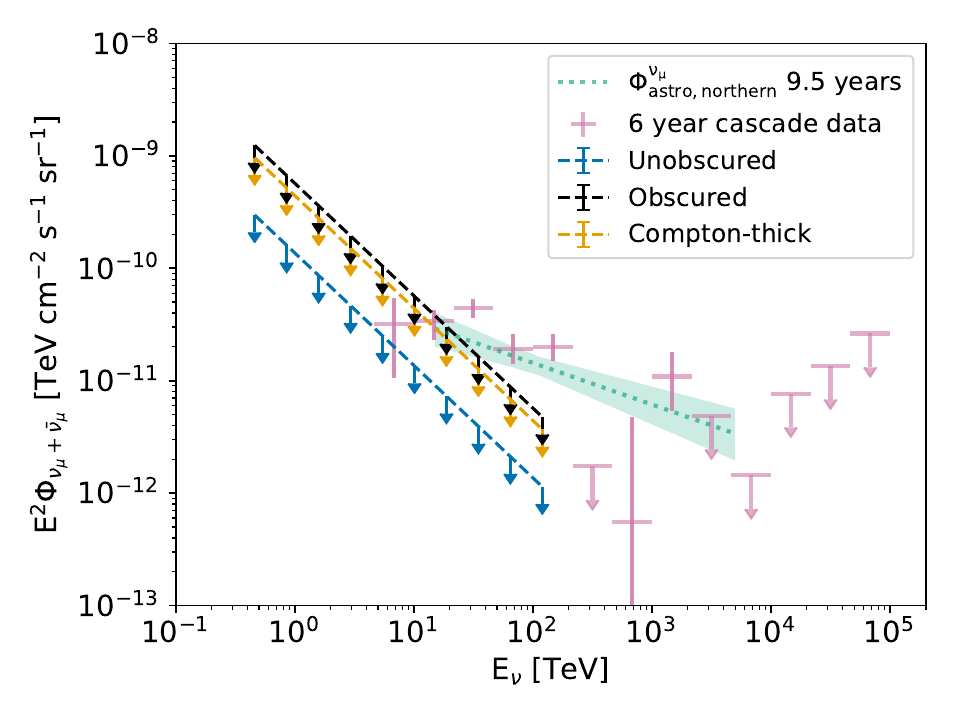}{0.42\linewidth}{}
          }
          
\caption{ 90\% CL flux upper limit for Left: all AGN in the catalog, blazars and non-blazars, and Right: for unobscured, obscured and Compton-thick sources. Flux upper limits are shown here for a spectral index of 3.0 since most samples result in a best-fit value of $\sim 3$.  These flux upper limits are displayed within the most relevant energy range for the analysis and are scaled using the catalog completeness factor described in Appendix~\ref{sec:appendix_catalog_correction}. For reference, the all-sky diffuse flux from muon tracks (light blue dotted line) \citep{Abbasi_2022} and from cascades (purple) \citep{Aartsen_2019} are shown on the plot. 
\label{fig:stacking_uppelimit_3.0}}
\end{figure}

Since there is no significant evidence of neutrino emission, we derive 90\% confidence level (CL) flux upper limits per flavor for power-law spectra with three different spectral indices $E^{-3.0}$, $E^{-2.5}$ and $E^{-2.0}$ (see Table~\ref{tab:stacking_results}). For spectral indices of $\gamma$ = 3.0, 2.5 and 2.0, the analyses with all AGN are most sensitive in the energy range 0.5~TeV to 0.1~PeV, 2.5~TeV to 1.1~PeV and 6~TeV to 10~PeV, respectively (see \autoref{fig:stacking_uppelimit_3.0} and \autoref{fig:stacking_uppelimit_2.0_2.5}). We used these energy ranges for all AGN samples since they are expected to be comparable (see Appendix~\ref{sec:appendix_energy_range}). The fluxes are scaled by the catalog completeness factor to take into account all unresolved sources (see Appendix~\ref{sec:appendix_catalog_correction}). 

We constrain the maximum contribution from all hard X-ray blazars to the astrophysical diffuse flux in \cite{Abbasi_2022} to be less than 7\%. This is evaluated using the flux upper limits obtained for blazars considering a power-law spectrum of $E^{-2.5}$ at 100 TeV and corrected for catalog completeness. Notably, the completeness factor for the blazars has large uncertainties as given in~Table~\ref{tab:catalog_completeness}. Previously, another IceCube result~\citep{2017ApJ...835...45A} constrained the contribution from \textit{Fermi}-2LAC blazars to be 27\%. There exists only a small overlap between the blazars in this analysis and the previous study. They independently conclude that blazars emit a small fraction of neutrinos. The flux upper limit obtained for all AGN, or non-blazar AGN considering a power law spectrum with an index $\gamma = 3.0$ at an energy $\sim30$~TeV is comparable to the diffuse flux. We cannot thus exclude that a major fraction of the diffuse flux can be attributed to these source classes.

\input{stacking_results}

\vspace{1em}
\textit{Individual source search} - From the list of 43 sources, the two sources with pre-trials significance > 3$\sigma$ are Seyfert galaxies NGC~1068 and NGC~4151 and their results are listed in Table \ref{tab:catalog_search_results}. NGC~1068 is found at a flux and spectral index consistent with previous results~\citep{doi:10.1126/science.abg3395}. Since it was observed with a high significance, we exclude it from the estimation of post-trials  $p$-value, evaluated as $1-(1-p_{\text{local}}$)$^N$ where $N$ is the number of sources. NGC~4151, the source with the highest FOM among the 43 AGN used for the individual source search is found to have a post-trials $p$-value of 1.67$\times$10$^{-3}$ (2.9$\sigma$). An excess of neutrinos was previously reported from a direction ($\sim 0.18^{\circ}$ angular deviation) compatible with NGC~4151 in \cite{doi:10.1126/science.abg3395}.

\begin{figure}[ht!]
\gridline{\fig{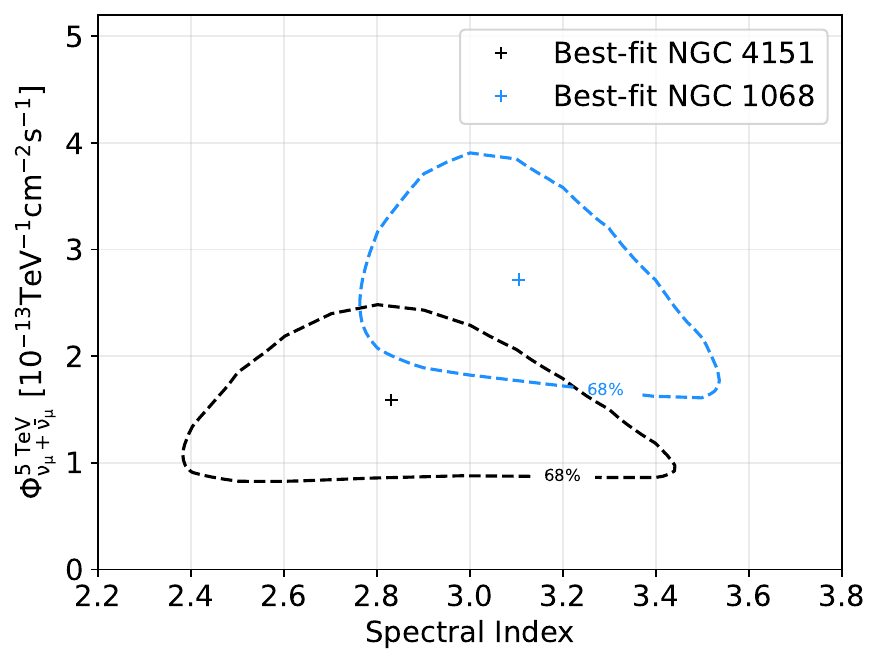}{0.42\linewidth}{}
          \fig{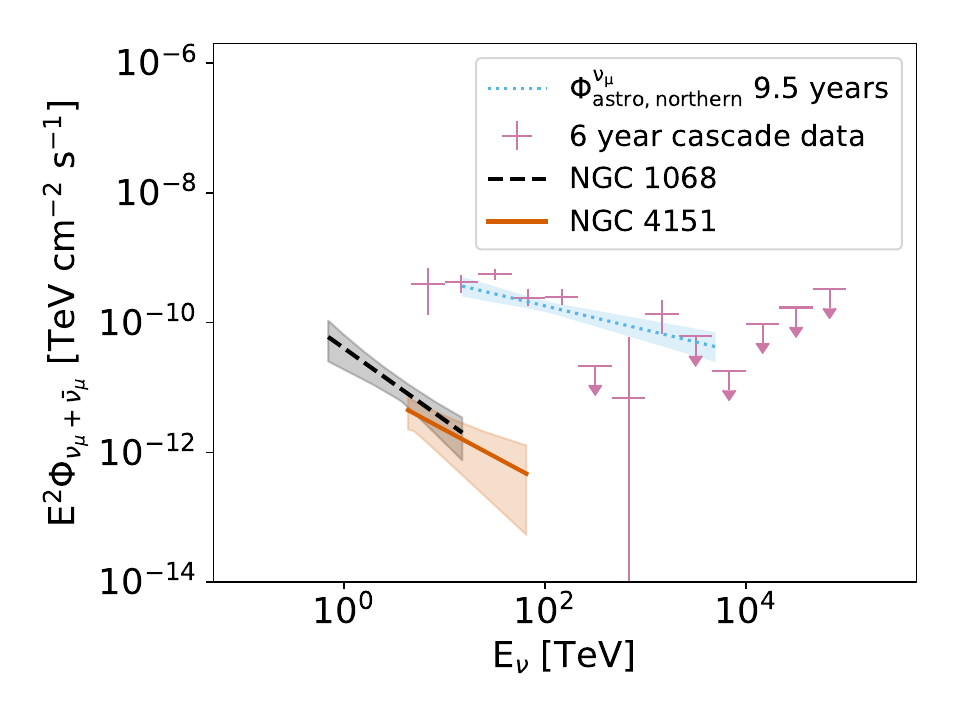}{0.42\linewidth}{}
          }
\caption{Left: 68\% CL contour lines for NGC~1068 (blue) and NGC~4151 (black) along with the best-fit values (crosses) of flux normalization and spectral index. Right: Flux as a function of energy from  NGC~1068 (black) and NGC~4151 (orange) obtained from the current analysis as compared to that of the all-sky total astrophysical neutrino flux from~\citealp{Abbasi_2022} showing 9.5 years of Northern tracks data and from~\citealp{PhysRevLett.125.121104} showing six years of cascade data. Lines are shown for the best-fit values and the shaded region represents the 68\% CL region which represents the statistical uncertainties. The systematic uncertainties are subdominant for this analysis. The energy range corresponds to the majority of the contribution towards the neutrino excess: we found the central 68\% of the contribution to the test statistic to be from neutrinos with energies of 0.7~TeV to 15~TeV for NGC~1068 and 4.3~TeV to 65.2~TeV for NGC~4151.}
\label{fig:flux_likelihood}
\end{figure}

While we do not reject the background hypothesis at > 3$\sigma$ significance, we can consider the excess to be astrophysical in origin and derive a best-fit flux normalization. \autoref{fig:flux_likelihood} (left) shows the best-fit values and 68\% CL contours from a likelihood scan using Wilk's theorem.  The contours are evaluated at 5~TeV where the correlation between fit parameters is minimum. The power-law spectra and uncertainties are shown in \autoref{fig:flux_likelihood} (right). The uncertainties are primarily statistical and the systematic uncertainties are subdominant as evident from the estimates found in multiple searches of neutrino sources, e.g.,~\cite{IceCube:2014stg, IceCube:2016tpw}.

\begin{figure}[ht!]
\begin{tabular}{cc}
\textbf{NGC~1068} & \textbf{NGC~4151} \\ \includegraphics[width=.4\textwidth]{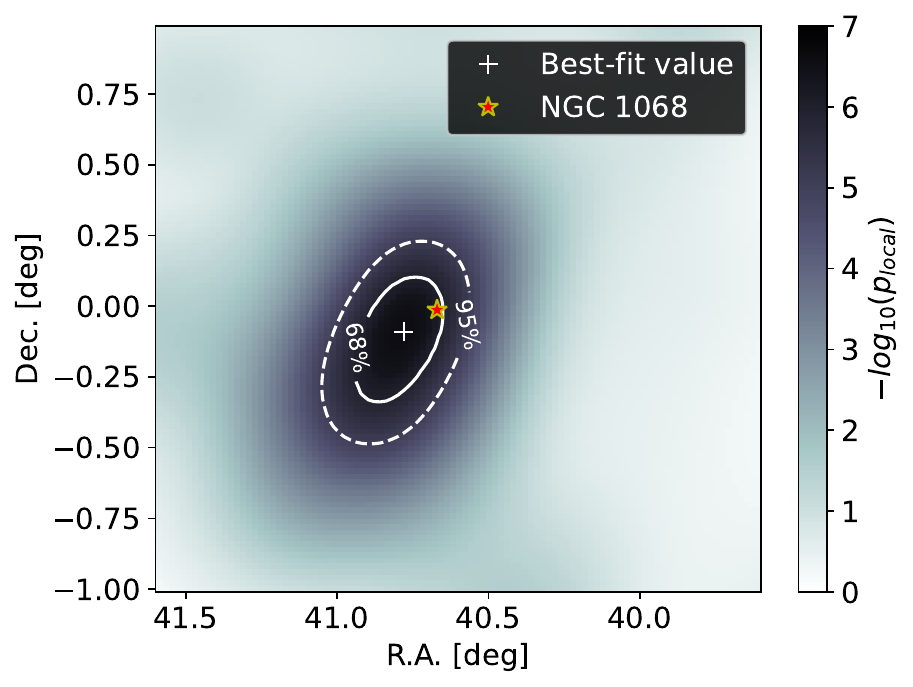} & 
\includegraphics[width=.4\textwidth]{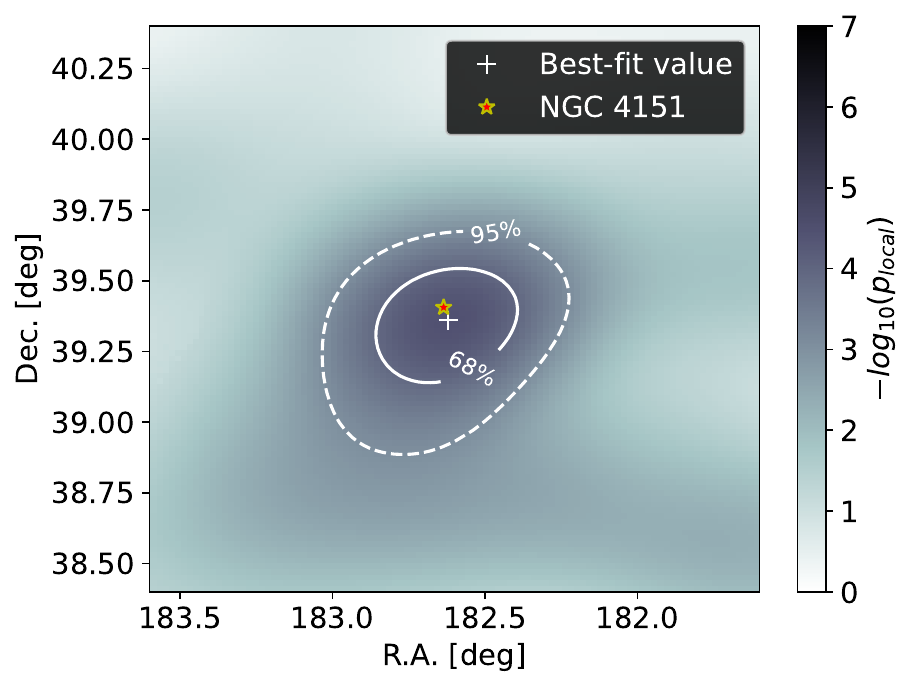} \\
(a) & (b) \\
\includegraphics[width=.4\textwidth]{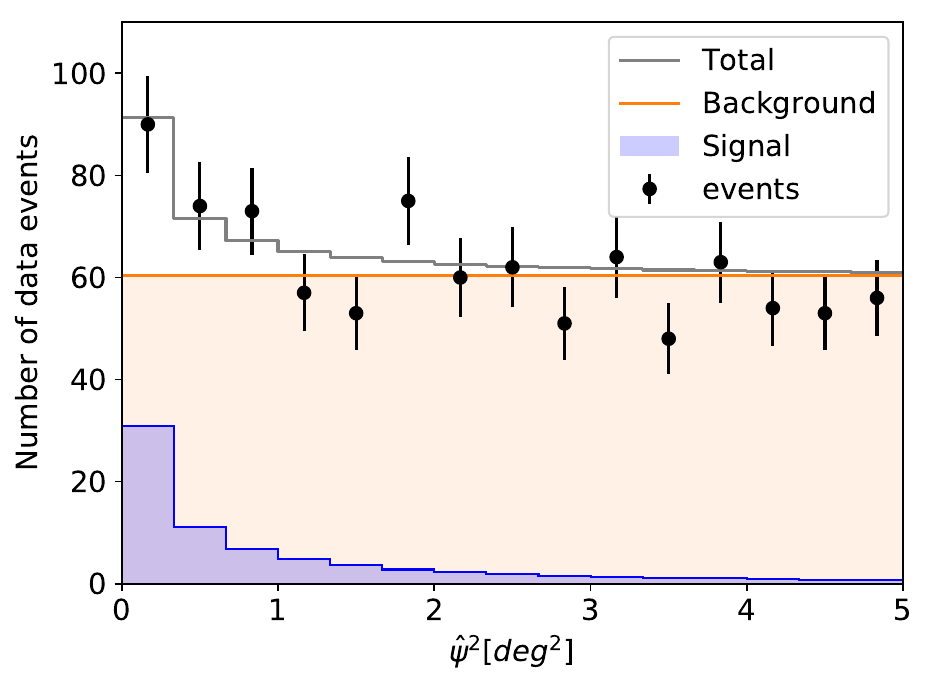} & 
\includegraphics[width=.4\textwidth]{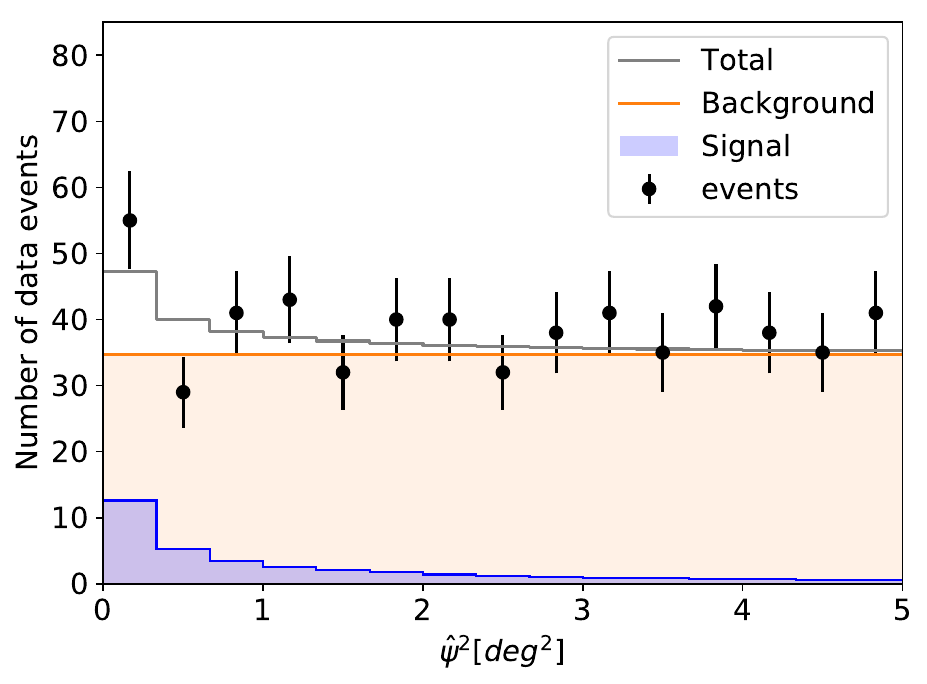} \\ 
(c) & (d) \\
\end{tabular}
\caption{Top row: High-resolution scan of the sky around the region of the two most significant source positions (a) NGC~1068 and (b) NGC~4151. For details of the method used to obtain the scans, see~\cite{doi:10.1126/science.abg3395}. The white cross shows the best-fit position and the red star shows the source position cataloged in BASS. The solid and dashed contours show the 68\% and 95\% CL region around the sources, respectively. Bottom row: Number of events as a function of the square of the angular distances from the source for (c) NGC~1068 and (d) NGC~4151. The best-fit astrophysical neutrino signal, the background, and their sum are shown in purple, orange, and grey, respectively. They are obtained using Monte Carlo simulations. The neutrino event data points are shown in black with error bars.}
\label{fig:scan_sources}
\end{figure}

We scan the region around the most significant sources as shown in \autoref{fig:scan_sources} (top). It shows the offset of the best-fit positions as a result of the analysis from the cataloged coordinates of the sources. The known source position lies within the 68\% contour line for both sources. See \autoref{fig:scan_sources} (bottom) for the distribution of the neutrino events with increasing angular distance from the sources.

\input{catalog_search_resuts}

\section{Discussion}\label{sec:discussion}

The most significant neutrino sources observed are two Seyfert galaxies, NGC~1068  and NGC~4151. However, the stacking analysis of non-blazar AGN (731 of 732 non-blazar AGN are Seyfert galaxies) shows no significant emission. This indicates that the potential neutrino emission is not directly proportional to the hard X-ray flux.

NGC~1068 and NGC~4151 are two nearby AGN, with photometric distances 11.14 $\pm$ 0.54 Mpc, 14.20 $\pm$ 0.88 Mpc \citep{2021AstBu..76..255T}, that are obscured ($N_{\text{H}}$ = $10^{24.95}$ cm$^{-2}$, $10^{22.71}$ cm$^{-2}$) and bright in X-rays (flux = 2.06$\times$10$^{-10}$ ergs cm$^{-2}$ s$^{-1}$, 5.26$\times$10$^{-10}$ ergs cm$^{-2}$ s$^{-1}$). They have similar SMBH masses, 1.3$\times$10$^7$ $M_{\odot}$ \citep{10.1093/mnras/staa1985} and 1.66$\times$10$^7$ $M_{\odot}$ \citep{Bentz_2022}. However, while NGC~1068 is a starburst galaxy, NGC~4151 shows little evidence of star formation \citep{10.1046/j.1365-8711.1999.02331.x}. Both sources show evidence of jet-disk interaction \citep{refId0,10.1093/mnras/staa1152}. The relativistic jets and the accretion disk are the two most promising sites for efficient particle acceleration, essential for neutrino production.

Acceleration mechanisms describing neutrino production predict a comparable flux of gamma rays and neutrinos. However, NGC~1068 is ``hidden'' in gamma rays as the neutrino flux exceeds the gamma-ray flux (see \autoref{fig:EM_nu_sed}). NGC~4151 is also an obscured source with the core surrounded by heavy amounts of dust and gas but there are no available gamma-ray observations. Neutrino emission from these sources was also tested in a complementary analysis of Seyfert galaxies \citep{IceCube:2024}. The two analyses overlap in the sources examined.
Nevertheless, the hypotheses tested, the neutrino dataset, and the analysis techniques are different. The complementary analysis tests specific models of emission~\citep{Kheirandish:2021wkm} using a dataset restricted to the northern sky. The conclusions from the two studies are consistent.

After the present work was unblinded, and while the manuscript was in preparation, the results of~\cite{PhysRevLett.132.101002} were shared with us and subsequently published.  In that work, the authors find a~$\sim 3\sigma$ neutrino signal from a population of Seyfert galaxies. The population of Seyferts they considered includes NGC 4151 and NGC 3079, the second and third most significant neutrino sources observed in the second analysis of this work (see Table~\ref{tab:catalog_search_upperlimits}).

The best-fit spectra found for the two sources are $\sim$$E^{-3}$ and softer than the diffuse astrophysical flux obtained by IceCube studies~\citep{Aartsen_2019, Abbasi_2022}. To resolve this discrepancy, we require either a dominant source population with a harder spectrum, or a class of similar sources with some emitting at higher energies than the observed ones. Considering NGC~1068 and NGC~4151 as two sources representative of a class of neutrino emitters, we find that they contribute $\sim 1\%$ of neutrinos to the diffuse flux but only $\sim 0.1\%$ to the total hard X-ray flux at a spectral index $\gamma = 3.0$. For the neutrino flux, we chose 15 TeV to evaluate the flux since this is within the sensitive energy range of both sources and the diffuse flux. We also assumed a direct relation between the hard X-ray and neutrino flux and corrected for catalog completeness. This suggests underlying physical parameters besides hard X-ray flux may determine neutrino emission despite the limitations of considering a sample of only two nearby AGN. The intrinsic fluxes from obscured sources, especially Compton-thick AGN, have high uncertainties as it is challenging to observe the emission from the core. In conclusion, growing evidence points to Seyfert galaxies as neutrino emitters. Better X-ray flux measurements of AGN especially for Compton-thick AGN may prove important to the discovery of high-energy neutrino sources.

\section*{Acknowledgements}
The IceCube collaboration acknowledges significant contributions to this manuscript from James DeLaunay, Sreetama Goswami and George C. Privon.

The authors gratefully acknowledge the support from the following agencies and institutions:
USA {\textendash} U.S. National Science Foundation-Office of Polar Programs,
U.S. National Science Foundation-Physics Division,
U.S. National Science Foundation-EPSCoR,
U.S. National Science Foundation-Office of Advanced Cyberinfrastructure,
Wisconsin Alumni Research Foundation,
Center for High Throughput Computing (CHTC) at the University of Wisconsin{\textendash}Madison,
Open Science Grid (OSG),
Partnership to Advance Throughput Computing (PATh),
Advanced Cyberinfrastructure Coordination Ecosystem: Services {\&} Support (ACCESS),
Frontera computing project at the Texas Advanced Computing Center,
U.S. Department of Energy-National Energy Research Scientific Computing Center,
Particle astrophysics research computing center at the University of Maryland,
Institute for Cyber-Enabled Research at Michigan State University,
Astroparticle physics computational facility at Marquette University,
NVIDIA Corporation,
and Google Cloud Platform;
Belgium {\textendash} Funds for Scientific Research (FRS-FNRS and FWO),
FWO Odysseus and Big Science programmes,
and Belgian Federal Science Policy Office (Belspo);
Germany {\textendash} Bundesministerium f{\"u}r Bildung und Forschung (BMBF),
Deutsche Forschungsgemeinschaft (DFG),
Helmholtz Alliance for Astroparticle Physics (HAP),
Initiative and Networking Fund of the Helmholtz Association,
Deutsches Elektronen Synchrotron (DESY),
and High Performance Computing cluster of the RWTH Aachen;
Sweden {\textendash} Swedish Research Council,
Swedish Polar Research Secretariat,
Swedish National Infrastructure for Computing (SNIC),
and Knut and Alice Wallenberg Foundation;
European Union {\textendash} EGI Advanced Computing for research;
Australia {\textendash} Australian Research Council;
Canada {\textendash} Natural Sciences and Engineering Research Council of Canada,
Calcul Qu{\'e}bec, Compute Ontario, Canada Foundation for Innovation, WestGrid, and Digital Research Alliance of Canada;
Denmark {\textendash} Villum Fonden, Carlsberg Foundation, and European Commission;
New Zealand {\textendash} Marsden Fund;
Japan {\textendash} Japan Society for Promotion of Science (JSPS)
and Institute for Global Prominent Research (IGPR) of Chiba University;
Korea {\textendash} National Research Foundation of Korea (NRF);
Switzerland {\textendash} Swiss National Science Foundation (SNSF).

\bibliography{references}{}

\bibliographystyle{aasjournal}

\section{Appendix }\label{sec:appendix}
\subsection{Neutrino Dataset}\label{sec:appendix_dataset}

Table~\ref{tab:Dataset_info} shows the different configurations in which the IceCube detector operated, the livetime in days and the number of events for each detector configuration. For the incomplete detector configurations with 40, 59 and 79 strings, the detector operated for about a year collecting data. With the complete configuration of 86 strings, there are various data-taking periods also about a year in length, each of which is called a ``season''. The table includes the breakdown of the IC 86 configuration into different seasons with the livetime and the number of events in each season.

\input{neutrino_dataset}

\subsection{Hard X-ray Sources}\label{sec:appendix_sources}  

The intrinsic X-ray flux is the observed flux after correcting for absorption and performing $k$-correction, i.e., shifting the flux such that it is in the 14~-~195~keV range in the source frame. The absorption arises mainly from photoelectric absorption and Compton scattering, which are both taken into account. The intrinsic X-ray flux and other parameters have been estimated in BASS using X-ray spectral analysis. Many models were tested by the authors to find the one that is most appropriate for each source.  Different models are used for the analysis of blazars and non-blazar AGN~\citep{1995MNRAS.273..837M}. The authors computed the flux in the energy range 14~-~195~keV by extrapolating the fluxes in the 14 - 150 keV energy range since the \textit{Swift}-BAT detector response for the highest energy range, 150~-~195~keV, is poor with signal to noise ratio $\sim100 - 1000$ times lower than other energy bands.

To evaluate the column density, the authors of the BASS catalog first used a simple power-law spectrum to fit the observations and then increased the number of free parameters as and when required.  From the spectral analysis, if a source resulted in column density, $N_{\text{H}}>10^{24}~\text{cm}^{-2}$, it is analyzed using a spheroidal torus model from \citealp{10.1111/j.1365-2966.2011.18207.x} and classified as Compton-thick in the BASS catalog~\citep{Ricci_2015}. 

\begin{figure}[h!]
    \centering
    \raisebox{1cm}{
    \includegraphics[width=0.52\textwidth]{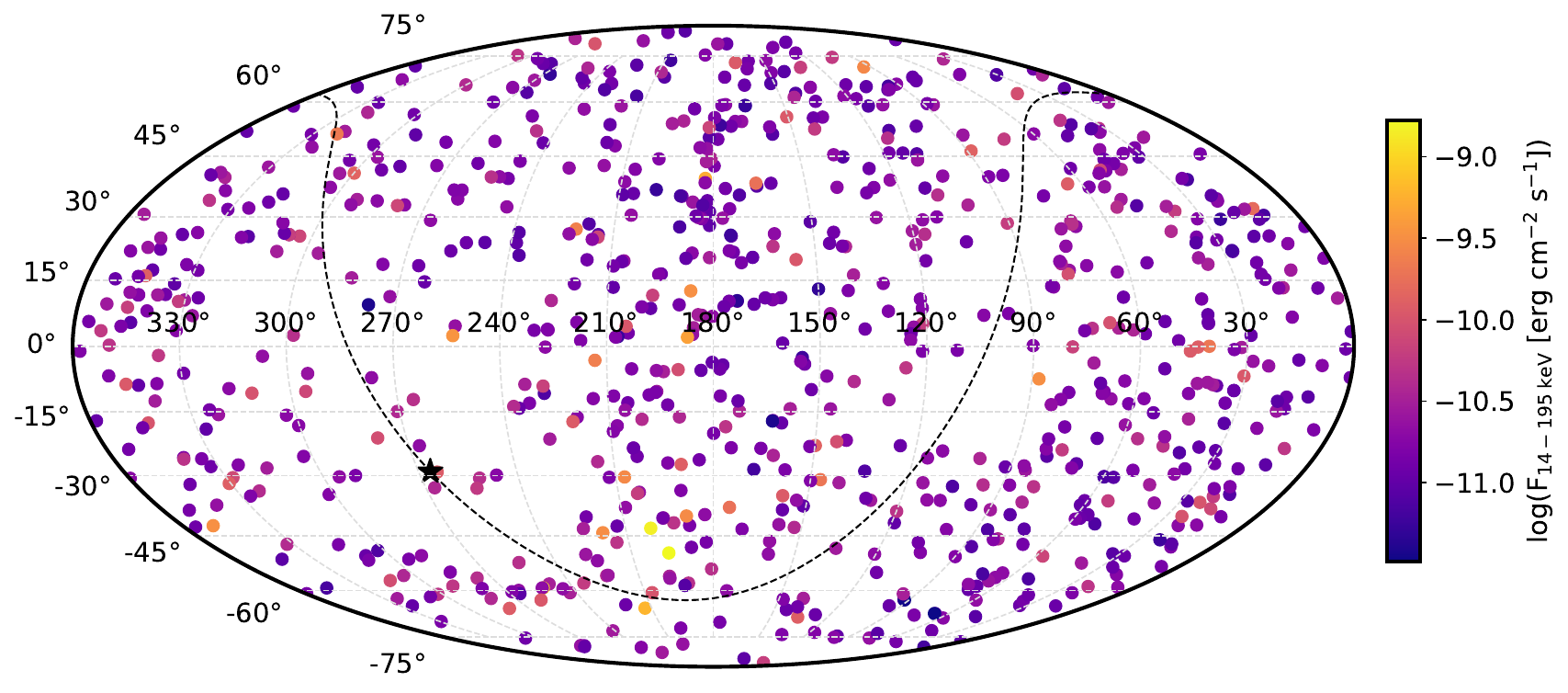}}
    \hfill
    \includegraphics[width=0.46\textwidth]{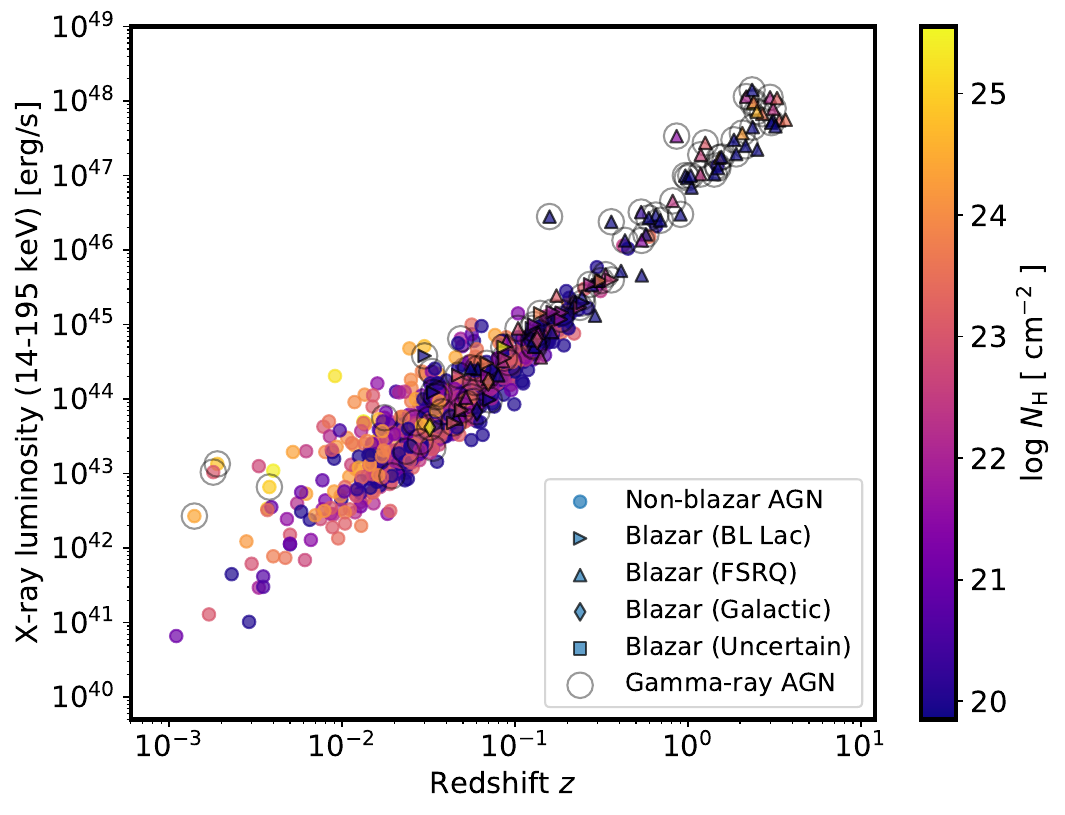}
    \caption{Left: A skymap in equatorial coordinates that shows all the hard X-ray AGN in the catalog used to perform the analyses in this study. The color scheme of the sources shows the hard X-ray intrinsic flux (in the 14 - 195 keV energy range). The black dashed line represents the Galactic plane and the star on the lower left represents the Galactic center.  Right: The distribution of the hard X-ray sources in the parameter space of luminosity as a function of redshift. The color of each data point represents the column density of the sources. The sources are shown using different markers that correspond to the type of AGN and the circled sources represent the AGN that have been detected in gamma rays and are found in the \textit{Fermi} 4-FGL catalog \citep{2020ApJS..247...33A}. The Galactic blazar is defined as a blazar with an additional component from the galaxy \citep{Koss_2022_cat}.}
    \label{fig:source_dist}
\end{figure}

In \autoref{fig:source_dist} (left), we show the positions of the hard X-ray AGN from the BASS catalog. In \autoref{fig:source_dist} (right), we can see from the y-axis that the sources in the catalog have a wide range of luminosity spanning around seven orders of magnitude. The most distant sources observed are around $z \sim$ 3 and are also the most luminous and as the distance increases, the AGN detected are more likely to be blazars with a higher luminosity and also detected in gamma rays. There are fewer non-blazar AGN observed at higher redshift in comparison since the dust and gas in the torus prevent any light from the core from being visible at very large distances. However, for the blazars their orientation makes the powerful jets point towards us and due to relativistic beaming, the emission from the jet is boosted, making them visible at a far greater distance. The jet is also able to penetrate any gas or dust it encounters even though the obscuration arises mostly from the torus whose axis aligns with the jet and hence, does not block any light from it.

\subsection{Hypothesis Testing}\label{sec:appendix_ts}

The statistical method followed to perform the two analyses is a hypothesis test. We evaluate the test statistic (TS) using the likelihood function given in Equation \ref{eq:Likelihood} and defined as follows, with a larger TS indicating that the data favor the signal hypothesis more than a lower TS:

\begin{equation}
\mathcal{TS} = -2 \log \left(\frac{\mathcal{L}(n_s=0)}
{\mathcal{L}(n_s=\hat{n}_s,\gamma=\hat{\gamma})} \right)  .  
\end{equation}

The pre-trials $p$-value, $p_{local}$, is defined as the survival probability of the observed TS value compared to a distribution of TS values computed by generating background-only pseudo-experiments. The background PDF is obtained by scrambling only the RA of neutrino events to preserve the declination distribution of the events. Each scrambling realization is a trial. We plot the TS values of these trials to obtain the background PDF. Then fit a $\chi^{2}_{\nu}$ distribution plus a delta function at TS = 0 to our histogram. 
To compute a $p$-value associated with a certain TS, we find the background-only TS distribution by performing repeated searches on events with randomized RA values. $p$-values are then converted to one-sided statistical significance.

\subsection{Stacked Search Methods and Additional Results}\label{sec:appendix_stacking}

\subsubsection{Catalog completeness correction}
\label{sec:appendix_catalog_correction}

To calculate the catalog completeness factor, we compute the fraction of the neutrino flux expected from the entire source population due to the sources contained in the catalog.

In the analysis, we hypothesize
\begin{equation}
    \frac{dN_{\nu}}{dE_{\nu}} \propto F_{\text{X-ray}}^{\text{int}}
\end{equation}

where $\frac{dN_{\nu}}{dE_{\nu}}$ is the neutrino differential energy spectrum and $F_{\text{X-ray}}^{\text{int}}$ is the intrinsic X-ray flux. 

Most of the sources in the BASS catalog are at relatively low redshift ($z_{\text{mean}}$ = 0.16) but for the most distant AGN, cosmological redshift plays a significant role in the observed neutrino flux. The flux is corrected for absorption and $k$-corrected. We measure the neutrino flux at a specific energy of 1 TeV and need to correct it to evaluate the intrinsic neutrino flux at the source. The observed neutrino differential energy spectrum is assumed to follow a power law  given by
\begin{equation}
    \frac{dN_{\nu}}{dE_{\nu}}^{\text{obs}} =  \frac{dN_{\nu}}{dE_{\nu}}^{\text{source}} (\text{1+z})^{-\gamma} \propto F_{\text{X-ray}}^{\text{int}}(\text{1+z})^{-\gamma}
\label{catcomp_prop}
\end{equation}

The values for $F_{\text{X-ray}}^{\text{int}}$  and redshift $z$ are available in the BASS catalog. Finally, by integrating the luminosity function of the source class (from~\citealp{BlazarCatComp} for blazars and from~\citealp{2014ApJ...786..104U} for non-blazar AGN), we obtain the contribution from the entire source population. The catalog completeness correction factor, or completeness factor in short, derived for each source class and spectral index are reported in Table \ref{tab:catalog_completeness}. The neutrino flux upper limits obtained from the stacked search for various spectral indices are divided by these values.

We need the sum of the neutrino signal expectation from both, the sources being tested and from the entire source class population in the observable universe. To calculate the completeness factor, we define a quantity $S(\gamma)$ using Equation~\ref{catcomp_prop} as $S(\gamma) = F_{X-ray}^{int} (1+z)^{-\gamma}$ and is proportional to the neutrino signal expectation. To find $S(\gamma)_{total}$, the sum of $S(\gamma)$ from the entire source class population, we use 

\begin{equation}
    S(\gamma)_{total} = \int^{4\pi}_0 \int^{z_{max}}_{0} \int^{L_{max}}_{L_{min}} S(L,z,\gamma) \phi (L,z)  \frac{dV}{dzd\Omega} dL dz d\Omega
\end{equation}
where $\phi(L,z)$ is the luminosity function and $\frac{dV}{dzd\Omega}$ is the co-moving volume per unit redshift and solid angle. To compute this integral, we need an accurate luminosity function that captures the source density through all relevant luminosities and redshifts. 

For the non-blazar AGN and their column density subclasses, we use the best-fit luminosity function from \citealp{2014ApJ...786..104U}. For their fit, the authors used X-ray surveys from multiple instruments that covered a combined energy range of 0.5~-~195~keV. The luminosity function is presented in the 2~-~10~keV range, so we use their spectral template of a cutoff power law to convert it to the 14~-~195~keV range. Since the luminosity function is in terms of intrinsic and $k$-corrected luminosity, no other adjustments need to be made.  

For the blazar AGN, we use the best-fit luminosity function from \citealp{BlazarCatComp}. They tested multiple models, so we used the model that has the lowest Akaike information criterion. The authors gave the luminosity function in terms of intrinsic and $k$-corrected luminosity, so no adjustments need to be made. 

When integrating these luminosity functions we use the same minimum and maximum luminosities as the authors; $L_{min} = 10^{43}$~erg~s$^{-1}$ and $L_{max} = 10^{50}$~erg~s$^{-1}$ for the blazars and $L_{min} = 10^{41}$~~erg~s$^{-1}$ and $L_{max} = 10^{47}$~erg~s$^{-1}$ for the non-blazars. We use the same maximum redshift of $z_{max} = 6$ for both classes. 

To calculate $S(\gamma)_{total}$ for the different classes of column density, we use the non-blazar luminosity function separated into its contributions from each class. We also need the blazar contribution for each class, but the blazar luminosity function is not a function of column density. Most blazars have low column densities as the jet is not traveling through the obscuring torus, so we add the blazar luminosity function to the unobscured non-blazar AGN luminosity function to calculate the unobscured $S(\gamma)_{total}$. Some BASS blazars fall into the obscured class (12\% of total BASS blazars), either through truly having that much obscuring material or through poor model fits. To account for this we add 0.12 times the blazar luminosity function to the obscured non-blazar luminosity function to calculate the obscured $S(\gamma)_{total}$. The fractional blazar contribution to each class are rough estimates, but the blazar total flux is sub-dominant to the non-blazar AGN flux by $\sim$2 orders of magnitude so little difference is made by changing these fractional contributions.

\begin{figure}[h!]
    \centering
    \includegraphics[width=0.65\textwidth]{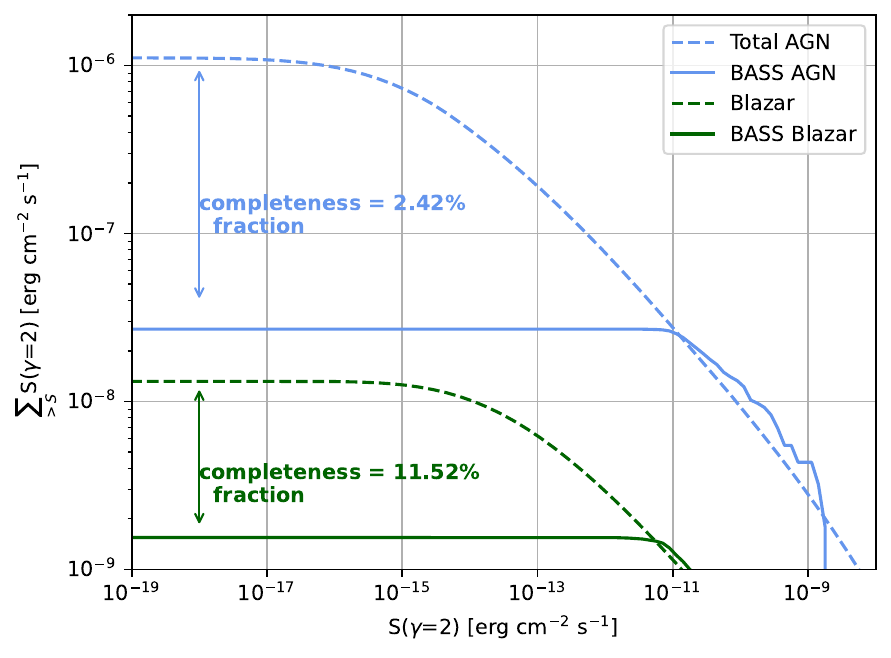}
    \caption{Plotted here are the inverse cumulative distributions of $S(\gamma)$ for $\gamma$ = 2, for both the entire population and the BASS sources. The distributions for both all AGN and just blazars are plotted. The catalog completeness fraction is also listed along with an arrow illustrating the difference between the numerator and denominator in the fraction. At the highest fluxes, there is an over fluctuation of bright sources which results in a higher value for sources in the catalog as compared to the total flux from all possible sources. Since the plot shows the cumulative distribution, this effect is not seen at the lower fluxes. }
    \label{fig:CatComp}
\end{figure}

In \autoref{fig:CatComp}, we have plotted the inverse cumulative $S(\gamma)$ for $\gamma$ = 2, as a function of $S(\gamma)$. The arrows in the figure show the difference between $S_{total}$ for just the sources in the catalog and the entire source class. The ratio of these two values gives our catalog completeness fraction. \autoref{fig:CatComp} outlines this for all AGN, which is dominated by non-blazar AGN, and for blazars. The curves for the BASS sources closely match the curves for the entire population up until the sensitivity of the catalog is reached. 

To find the error of $S(\gamma)_{total}$ we use a bootstrapping method along with the reported errors for each fit parameter from both luminosity functions. We randomly sample the parameters' values from their error PDFs, assuming they are Gaussian and uncorrelated. They are most likely correlated, but we do not have that information and the assumption of no correlation gives a larger and more conservative error. Then, we calculate $S(\gamma)_{total}$ for many samples and find the 16th and 84th percentiles of the distribution of $S(\gamma)_{total}$ to determine the 1$\sigma$ range of the completeness factors. The errors are shown in Table \ref{tab:catalog_completeness} along with the completeness factors for three different $\gamma$'s. 

\input{catalog_completeness}

\subsubsection{Trials Factor Correction}\label{sec:appendix_trial_correc}

\begin{figure}[h!]
    \centering
    \includegraphics[width=0.65\textwidth]{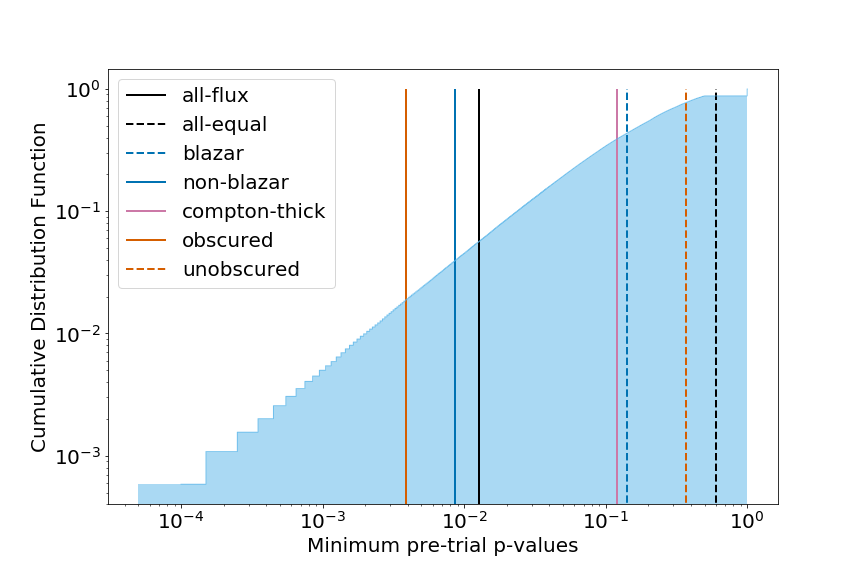}
    \caption{The cumulative distribution function of the minimum of the  $p$-values from each AGN sample resulting from background trials, and the pre-trials  $p$-values of the samples in the unblinded analysis. All-flux and all-equal refer to the samples containing all 836 AGN tested using hard X-ray flux as weights and equal weights, respectively.}
    \label{fig:trial_factor_correction}
\end{figure}

We have tested seven different hypotheses in the stacking analysis (see Section~\ref{subsec:analysis I}) using different sub-classes of AGN with a significant overlap in the source populations. To evaluate the correlation between the different AGN samples used to test each hypothesis we performed a large number of trials, $\mathcal{O}(10^{5})$, since a simple analytical function as was used previously for the individual source search is not sufficient in this case. From each background trial, we computed the  $p$-values for each of the seven AGN samples and found the minimum  $p$-value. \autoref{fig:trial_factor_correction} shows the distribution of the minimum  $p$-values. The corresponding value of $p_{\text{local}}$ gives the post-trials  $p$-value. For the AGN samples tested, the pre-trials $p$-values and pre-trial significances are reported in Table~\ref{tab:stacking_results}.

\subsubsection{Energy Range Evaluation}\label{sec:appendix_energy_range}

As the sensitivity of the IceCube detector is not constant for all neutrinos and varies with energy, we evaluate the energy range most relevant to the analyses. We follow two methods for the two analyses. This range reflects most of the neutrino events that contribute to the TS.

Since we find flux upper limits in the stacking analysis and the results do not show a high significance, we utilize pseudo-signal events from Monte Carlo instead of experimental data to determine the central 90\% energy range. To evaluate the upper bound of the energy range, we performed many trials by injecting signals over the background, with a fixed value of lower energy bound and varying the upper energy bound. Next, we compute the sensitivity for each energy range and plot them as a function of the energy. The lowest value of the higher bound energy for which the evaluated sensitivity is below 1.05 times the total sensitivity flux evaluated over an energy range of zero to infinity gives us the most sensitive higher bound energy value. To evaluate the lower bound of the energy range, we performed many trials by injecting signals over the background with a fixed value of higher energy bound and varying the lower energy bound to find the sensitivity for each energy range and plot them as a function of the energy. The lowest value of the varying lower bound energy for which the evaluated sensitivity is above 1.05 times the total sensitivity flux evaluated over an energy range of zero to infinity gives us the most sensitive higher bound energy value.

The most sensitive energy range is not expected to vary significantly with the different  AGN samples but will change for different spectral indices. Therefore, we evaluate the energy range using all AGN and use the values for the other samples as well. The evaluated energy range where the analysis is most sensitive for $\gamma = 2.0$ is 6~TeV to 10~PeV, $\gamma = 2.5$ is 2.5~TeV to 1.1~PeV and $\gamma = 3.0$ is  0.5~TeV to 0.1~PeV.

For the individual source search, the relevant energy ranges were determined using the experimental data collected from muon tracks over 12 years. The resultant fluxes for the two sources, NGC~1068 and NGC~4151, were obtained assuming a power-law spectrum. Even though these fluxes extend across the entire energy range covered by the neutrino dataset, the most relevant energy range corresponds to that which contributes 68\% towards the TS obtained for each source. 

In the case of NGC~1068, the energy range is 0.7 TeV to 15 TeV. This is slightly different from the value reported for NGC~1068 in \cite{doi:10.1126/science.abg3395}, where the energy range is 1.5 TeV to 15 TeV. This discrepancy is due to the different datasets being used for the two analyses and their different effective areas. While in this analysis, we used the Point Source Tracks sample which includes tracks from the whole sky (for details, see~\cite{PhysRevLett.124.051103}), the previous analyses used the Northern Tracks sample which focuses on muon tracks from the Northern sky (for details, see~\cite{aartsen2019search}).

\subsubsection{Stacked Search: Additional Results}\label{sec:appendix_stacking_res}
\begin{figure}[ht!]
\gridline{\fig{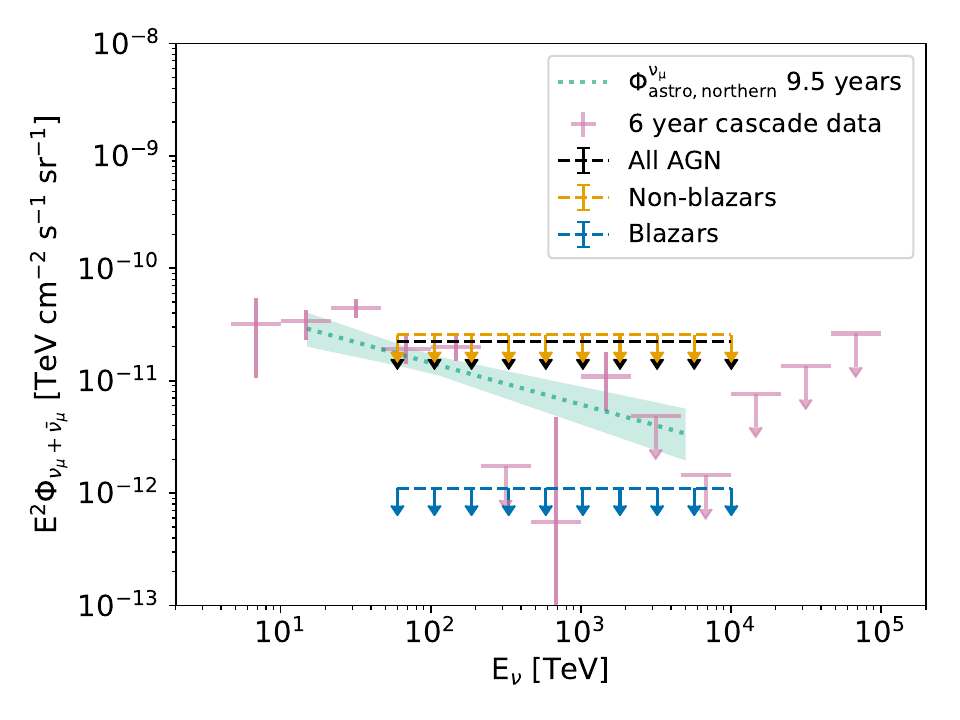}{0.42\linewidth}{(a)}
          \fig{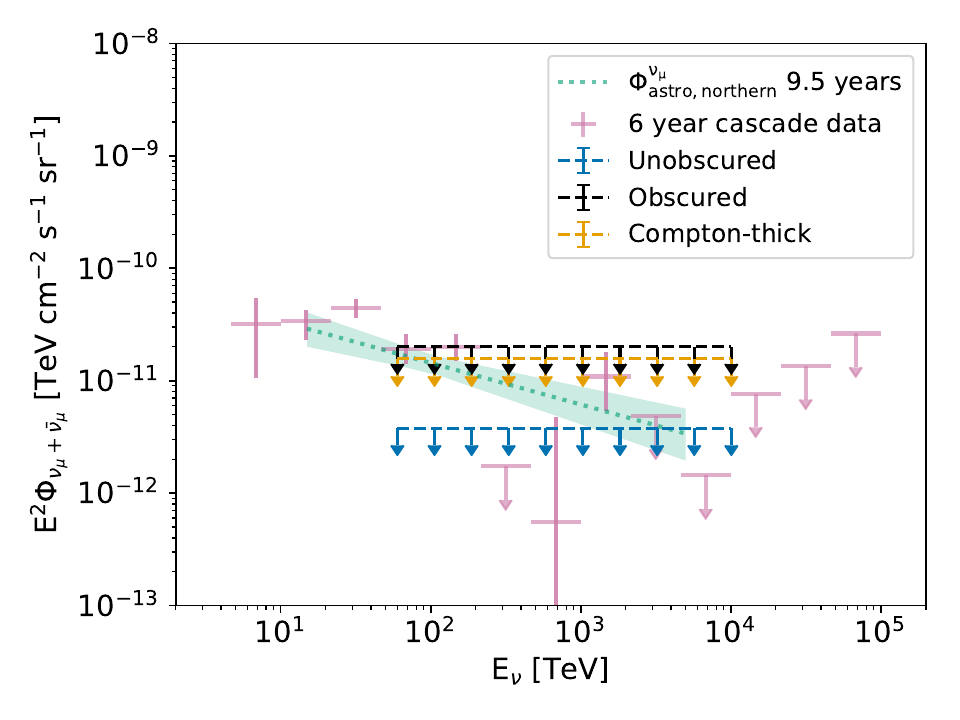}{0.42\linewidth}{(b)}
          }
\gridline{\fig{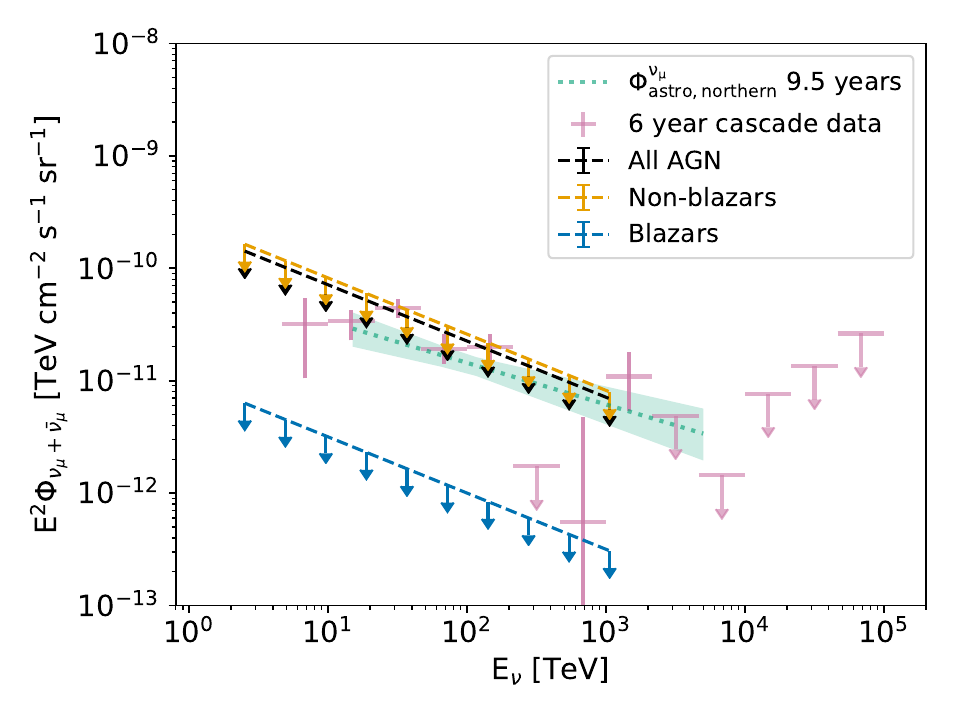}{0.42\linewidth}{(c)}
          \fig{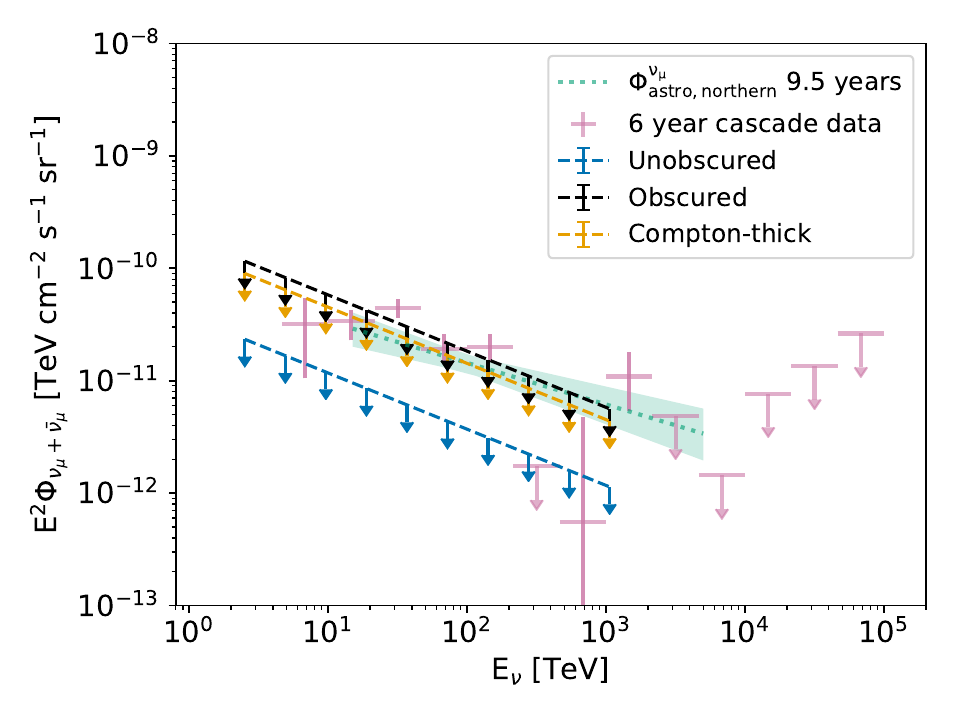}{0.42\linewidth}{(d)}
          }
\caption{90\% CL flux upper limit for (a) and (c): all AGN, AGN subclasses of blazars and non-blazars, and (b) and (d): unobscured, obscured and Compton-thick sources. The flux is for power-law spectra with a spectral index of 2.0 (top row) and 2.5 (bottom row). These flux values are derived using the completeness correction described in Appendix~\ref{sec:appendix_catalog_correction}. For reference, the all-sky diffuse flux from muon tracks \citep{Abbasi_2022} in light blue dotted line and cascades \citep{Aartsen_2019} in purple are shown on the plot.
\label{fig:stacking_uppelimit_2.0_2.5}}
\end{figure}

In addition to the flux upper limits evaluated for index $\gamma = 3.0$, we computed upper limits for indices, $\gamma = 2.5$  and $\gamma = 2.0$  as well. These flux upper limits are given in Table~\ref{tab:stacking_results} and shown in \autoref{fig:stacking_uppelimit_2.0_2.5}.

\subsection{Individual Source Search: Source Selection}\label{sec:appendix_point_source_selection}

\begin{figure}[h!]
\gridline{\fig{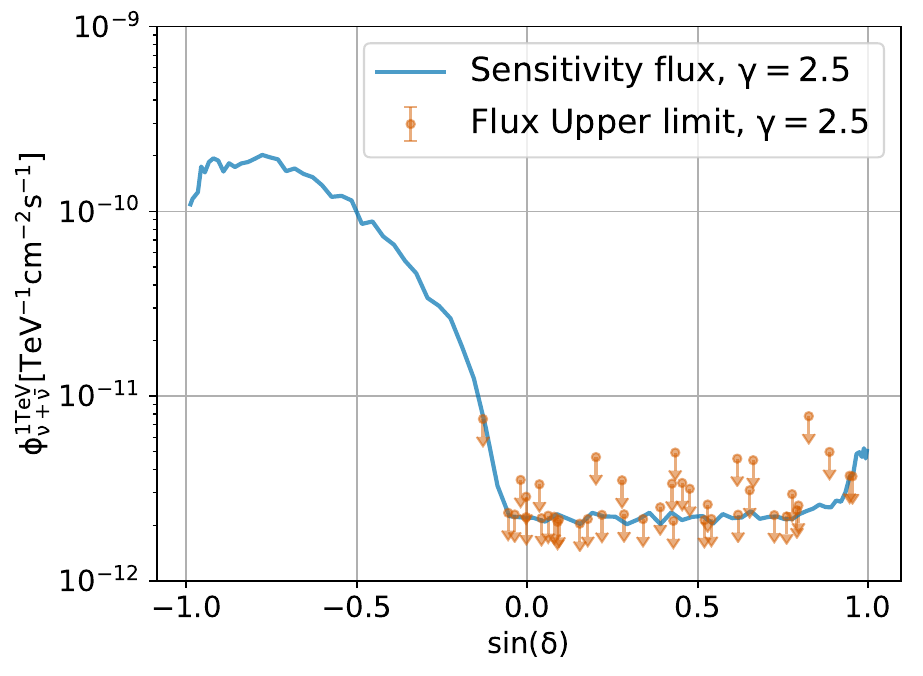}{0.42\textwidth}{}
          \fig{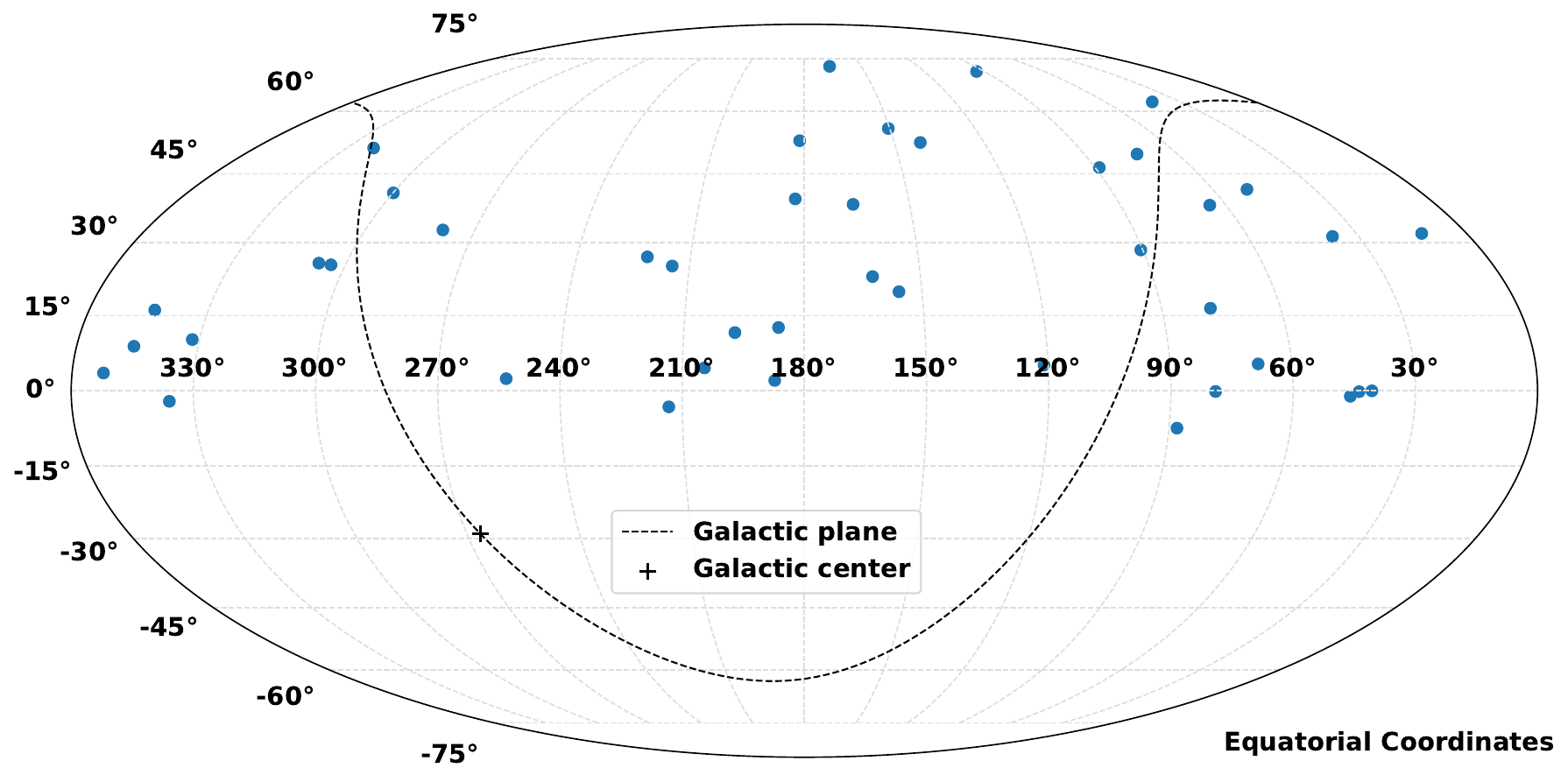}{0.52\textwidth}{}
          }
\caption{Left: Neutrino sensitivity flux (in blue) and the flux upper limits (in orange) at normalization energy 1 TeV and power-law spectral index $\gamma$ = 2.5 as a function of the sine of declination. Right: Skymap showing the 43 sources from which we search for the emission of neutrinos, individually. Most sources are in the northern hemisphere as seen in \autoref{fig:sens_n_skymap} (right).}
\label{fig:sens_n_skymap}
\end{figure}

To evaluate the ``figure of merit'' which formed the basis of the selection of 43 sources to be analyzed more closely in the individual source search in Section~\ref{subsec:analysis II},  we first computed the neutrino flux sensitivity, defined as the required signal, in terms of flux or number of neutrinos, that yields a $p$-value smaller than 0.5 in 90$\%$ of the trials,  as a function of the declination for spectral index $\gamma = 2.5$. The sensitivity allows the IceCube detector to observe a source at different declinations and is shown in \autoref{fig:sens_n_skymap} (left). The list's final selection of 43 sources and their distribution in the sky is shown in \autoref{fig:sens_n_skymap} (right).

\subsection{Individual Source Search: Additional Results}\label{sec:appendix_point_source_res}

\begin{figure}[h!]
    \centering
    \includegraphics[width=0.75\textwidth]{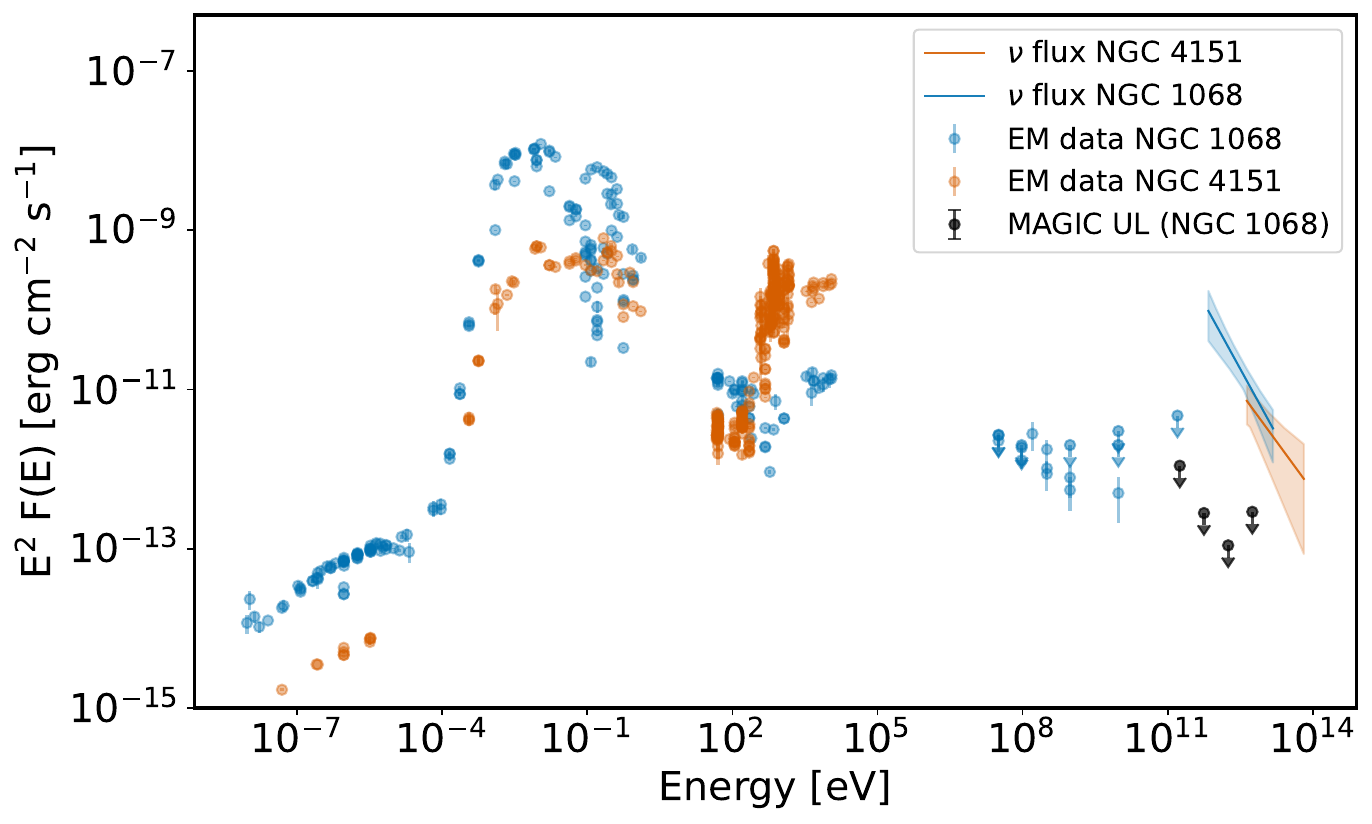}
    \caption{The spectral energy distribution of the electromagnetic (EM) emission from  NGC~4151 and NGC~1068 and their neutrino spectrum obtained from this analysis. Electromagnetic data and upper limits are taken from \url{https://tools.ssdc.asi.it/SED/}.}
    \label{fig:EM_nu_sed}
\end{figure}

We calculated the $p_{\text{local}}$ for each of the sources, which is reported in Table~\ref{tab:catalog_search_upperlimits}. For the two most significant sources, we show their electromagnetic spectra observed by different telescopes overlaid with the neutrino spectrum obtained from this analysis in \autoref{fig:EM_nu_sed}. 

Other than the two sources found with the highest significance, we derived the flux upper limits for each source at three different power-law indices $\gamma$ = 2.0, 2.5 and 3.0. See the flux upper limits in Table~\ref{tab:catalog_search_upperlimits}.

\input{source_list_upperlimit}

\end{document}

%% file: stacking_results.tex
\begin{deluxetable}{lccccccc}
\tablecaption{Summary of results from the stacked search.}
\label{tab:stacking_results}
\tablehead{\colhead{Sample}  & \colhead{No. of sources} 
& \colhead{$\hat{n}_{s}$}    & \colhead{$\hat{\gamma}$} 
& \colhead{p$_{local}$}        & \colhead{$\phi_{90\%}$} 
& \colhead{$\phi_{90\%}$}    & \colhead{$\phi_{90\%}$}\\
\colhead{}                   & \colhead{}         
& \colhead{}                 & \colhead{}           
& \colhead{(significance)}   & \colhead{($\gamma = 2.0$)} 
& \colhead{($\gamma = 2.5$)} & \colhead{($\gamma = 3.0$)} 
}
\decimals
\startdata
All AGN       & 836 & 161  & 2.89 &  0.01 (2.2$\sigma$) & 0.68  & 8.97  & 39.25 \\ 
Blazars       & 104 & 10   & 2.04 &  0.14 (1.1$\sigma$) & 0.16  & 1.83  & 7.38 \\
Non-blazars   & 732 & 180  & 3.02 &  0.01 (2.4$\sigma$) & 0.75  & 9.89  & 41.94 \\
Compton-thick & 56  & 45   & 3.14 &  0.12 (1.2$\sigma$) & 0.30  & 3.63  & 14.40 \\
Obscured      & 323 & 148  & 2.91 & 0.003 (2.7$\sigma$) & 0.57 & 6.85  & 27.06\\
Unobscured    & 457 & 0    & 0    &  1.00 (0.0$\sigma$) & 0.24  & 3.02  & 13.85
\enddata
\tablecomments{Flux upper limit normalized at 1 TeV is in units of $10^{-11}$ TeV$^{-1}$cm$^{-2}$s$^{-1}$. Best-fit number of neutrinos, $\hat{n}_{s}$ is the sum over all sources in the sample. } 
\end{deluxetable}

%% file: catalog_search_resuts.tex
\begin{deluxetable}{lcccccc}
\tablecaption{The two most significant sources from the individual source search.}
\label{tab:catalog_search_results}
\tablewidth{0pt}
\tablehead{\colhead{Name} & \colhead{Type} & \colhead{$\hat{n}_s$} & \colhead{$\hat{\gamma}$}  & \colhead{$\phi_{\nu_{\mu}+\bar{\nu}_{\mu}}^{\text{1~TeV}}$} & \colhead{$p_{local}$ } &  \colhead{post-trial $p$-value } \\
\colhead{} & \colhead{} & \colhead{}  & \colhead{}  &\colhead{(TeV$^{-1}$cm$^{-2}$s$^{-1}$)} &\colhead{(significance)} &\colhead{(significance)}
}
\startdata
 NGC 1068 & Compton-thick Seyfert  & 81.7 & 3.10$^{+0.26}_{-0.22}$ & 4.02$_{-1.52}^{+1.58}  \times10^{-11}$ & 1.27$\times10^{-6}$ (4.7$\sigma$) & -
\\
NGC 4151 &  Obscured Seyfert & 49.8 & 2.83$^{+0.35}_{-0.28}$ & 1.51$_{-0.81}^{+0.99}\times10^{-11}$ &  3.99$\times10^{-5}$ (3.9$\sigma$) & 1.67$\times10^{-3}$ (2.9$\sigma$)
\enddata
\end{deluxetable}

%% file: neutrino_dataset.tex
\begin{deluxetable}{ccc}
\tablecaption{Details of the neutrino dataset used for the analysis.}
\label{tab:Dataset_info}
\tablewidth{0pt}
\tablehead{
\colhead{Configuration} & \colhead{Livetime (days)}  & \colhead{No. of Events} 
}
\startdata
IC40       & 06 Apr 2008 - 20 May 2009 (376.4) &  36900 \\ 
IC59       & 20 May 2009 - 31 May 2010 (353.6) & 107011 \\
IC79       & 01 Jun 2010 - 13 May 2011 (316.0) & 93133 \\
\hline
IC86, 2011 & 13 May 2011 - 15 May 2012 (340.1) & 119169 \\
IC86, 2012 & 26 Apr 2012 - 02 May 2013 (327.7) & 116715 \\
IC86, 2013 & 18 Apr 2013 - 05 May 2014 (355.6) & 126337 \\
IC86, 2014 & 10 Apr 2014 - 18 May 2015 (365.5) & 129823 \\
IC86, 2015 & 24 Apr 2015 - 20 May 2016 (365.3) & 130434 \\
IC86, 2016 & 20 May 2016 - 18 May 2017 (357.2) & 126438 \\
IC86, 2017 & 18 May 2017 - 10 Jul 2018 (405.6) & 145602 \\
IC86, 2018 & 19 Jun 2018 - 17 Jul 2019 (362.5) & 129230 \\
IC86, 2019 & 28 Jun 2019 - 29 May 2020 (304.7) & 109616 \\
\hline
IC86, all  & 13 May 2011 - 29 May 2020 (3184.2) & 1133364 
\enddata
\tablecomments{The start date of some configurations is earlier than the end date of the previous season. This is because these dates denote the start date of test runs of the new processing and the data-taking begins once the previous run ends.}
\end{deluxetable}

%% file: catalog_completeness.tex
\capstartfalse
\begin{deluxetable}{lccc}
\tablecaption{Fractions showing the completeness factors of the source populations.}
\label{tab:catalog_completeness}
\tablewidth{0pt}
\tablehead{\colhead{Category} & \colhead{Completeness fraction} & \colhead{Completeness fraction}& \colhead{Completeness fraction}\\
\colhead{} & \colhead{$\gamma$=2.0 (E$^{-2.0}$)} & \colhead{$\gamma$=2.5 (E$^{-2.5}$)}& \colhead{$\gamma$=3.0 (E$^{-3.0}$)}
}
\startdata
All AGN          &  2.42$^{+0.59}_{-0.59}\%$  &  3.17$^{+0.72}_{-0.74}\%$   & 4.01$^{+0.92}_{-0.89}\%$\\
Non-blazar AGN   &  2.31$^{-0.62}_{-0.51}\%$  &  3.04$^{+0.75}_{-0.64}\%$   &  3.86$^{+0.95}_{-0.80}\%$\\
Blazars          & 11.52$^{+24.15}_{-8.74}\%$ & 14.58$^{+26.58}_{-10.86}\%$ &  17.97$^{+28.74}_{-12.50}\%$\\
Unobscured       &  5.05$^{+1.14}_{-1.57}\%$  &  6.49$^{+1.32}_{-1.89}\%$   &  8.06$^{+1.68}_{-2.23}\%$\\
Obscured         &  2.26$^{+0.59}_{-0.52}\%$  &  2.98$^{+0.71}_{-0.65}\%$   &  3.78$^{+0.90}_{-0.81}\%$\\
Compton-thick    &  1.52$^{+0.42}_{-0.34}\%$  &  2.03$^{+0.50}_{-0.43}\%$   &  2.59$^{+0.64}_{-0.54}\%$
\enddata
\end{deluxetable}
\capstarttrue

%% file: source_list_upperlimit.tex
\capstartfalse
\begin{deluxetable}{ccccccccccc}
    \tablecaption{Summary of results from the individual source search.}
    \label{tab:catalog_search_upperlimits}
\tablehead{\colhead{Type} & \colhead{Source} & \colhead{$\alpha$ [deg]} & \colhead{$\delta$ [deg]} & \colhead{$\hat{n}_s$} & \colhead{$\hat{\gamma}$} & \colhead{$p_{local}$} & \colhead{X-ray flux} & \colhead{$\phi^{2.0}_{90\%}$} & \colhead{$\phi^{2.5}_{90\%}$} & \colhead{$\phi^{3.0}_{90\%}$}}
\startdata
Seyfert & NGC1068 & 40.67 & -0.01 & 63.18 & 3.1 & 1.27$\times10^{-6}$ & 206.0 & - & - & - \\
 & NGC4151 & 182.64 & 39.41 & 43.67 & 2.83 & 3.99$\times10^{-5}$ & 525.7 & - & - & - \\
 & NGC3079 & 150.49 & 55.68 & 29.53 & 4.0 & 0.003 & 112.6 & 13.84 & 77.91 & 203.50 \\
 & MCG+4-48-2 & 307.15 & 25.73 & 21.74 & 3.75 & 0.060 & 63.9 & 6.44 & 49.50 & 150.52 \\
 & NGC4992 & 197.27 & 11.63 & 17.41 & 2.86 & 0.086 & 66.8 & 4.89 & 46.79 & 166.27 \\
 & 3C111 & 64.59 & 38.03 & 18.22 & 3.45 & 0.108 & 118.7 & 7.43 & 45.91 & 127.20 \\
 & NGC1275 & 49.95 & 41.51 & 16.13 & 3.55 & 0.117 & 79.2 & 7.42 & 45.01 & 123.17 \\
 & Q0241+622 & 41.24 & 62.47 & 18.4 & 3.84 & 0.129 & 89.6 & 10.17 & 49.92 & 120.86 \\
 & CygnusA & 299.87 & 40.73 & 1.84 & 1.38 & 0.141 & 137.9 & 5.98 & 30.95 & 114.64 \\
 & NGC1194 & 45.95 & -1.1 & 12.74 & 3.75 & 0.229 & 114.5 & 2.69 & 35.22 & 144.95 \\
 & NGC5548 & 214.5 & 25.14 & 11.18 & 4.0 & 0.246 & 75.9 & 4.86 & 33.63 & 103.11 \\
 & Z164-19 & 221.4 & 27.03 & 8.43 & 4.0 & 0.252 & 250.1 & 4.74 & 33.90 & 102.57 \\
 & IRAS05589+2828 & 90.54 & 28.47 & 7.58 & 4.0 & 0.268 & 62.4 & 4.89 & 31.53 & 90.33 \\
 & 4C+50.55 & 321.16 & 50.97 & 6.41 & 3.41 & 0.329 & 217.3 & 5.56 & 29.49 & 79.88 \\
 & Mrk348 & 12.2 & 31.96 & 5.27 & 4.0 & 0.356 & 159.2 & 4.38 & 25.92 & 81.12 \\
 & NGC1142 & 43.8 & -0.18 & 7.43 & 3.18 & 0.359 & 123.7 & 2.26 & 28.56 & 113.46 \\
 & Mrk417 & 162.38 & 22.96 & 1.72 & 2.44 & 0.407 & 50.0 & 3.46 & 25.03 & 106.33 \\
 & NGC4102 & 181.6 & 52.71 & 4.05 & 3.75 & 0.441 & 71.8 & 4.91 & 25.60 & 65.94 \\
 & Mrk1040 & 37.06 & 31.31 & 3.51 & 3.49 & 0.460 & 61.1 & 3.53 & 21.35 & 67.19 \\
 & Mrk110 & 141.3 & 52.29 & 0.41 & 1.43 & 0.466 & 57.1 & 4.79 & 24.10 & 64.50 \\
 & 3C120 & 68.3 & 5.35 & 4.69 & 3.82 & 0.472 & 93.4 & 2.19 & 21.49 & 78.00 \\
 & NGC5252 & 204.57 & 4.54 & 0.73 & 2.24 & 0.493 & 106.3 & 2.01 & 22.24 & 84.06 \\
 & NGC7469 & 345.82 & 8.87 & 0.17 & 2.58 & 0.534 & 64.3 & 2.16 & 20.42 & 73.51 \\
 & IRAS05078+1626 & 77.69 & 16.5 & 0.0 & 0.0 & 1.0 & 92.3 & 2.90 & 22.94 & 71.02 \\
 & Mrk1210 & 121.02 & 5.11 & 0.0 & 0.0 & 1.0 & 61.5 & 1.96 & 20.87 & 85.90 \\
 & NGC2110 & 88.05 & -7.46 & 0.0 & 0.0 & 1.0 & 317.5 & 3.51 & 75.21 & 608.59 \\
 & NGC7682 & 352.27 & 3.53 & 0.0 & 0.0 & 1.0 & 55.8 & 1.98 & 22.55 & 85.59 \\
 & Ark120 & 79.05 & -0.15 & 0.0 & 0.0 & 1.0 & 68.0 & 1.80 & 22.15 & 91.29 \\
 & 3C382 & 278.76 & 32.7 & 0.0 & 0.0 & 1.0 & 75.0 & 3.52 & 21.65 & 64.13 \\
 & 2MASXJ20145928+2523010 & 303.75 & 25.38 & 0.0 & 0.0 & 1.0 & 73.6 & 3.12 & 21.15 & 67.60 \\
 & NGC3516 & 166.7 & 72.57 & 0.0 & 0.0 & 1.0 & 113.9 & 8.01 & 36.84 & 93.62 \\
 & NGC6240 & 253.25 & 2.4 & 0.0 & 0.0 & 1.0 & 348.2 & 1.89 & 21.86 & 85.01 \\
 & NGC3227 & 155.88 & 19.87 & 0.0 & 0.0 & 1.0 & 106.6 & 2.90 & 21.62 & 69.19 \\
 & NGC5506 & 213.31 & -3.21 & 0.0 & 0.0 & 1.0 & 234.7 & 1.87 & 23.38 & 94.73 \\
 & UGC3374 & 88.72 & 46.44 & 0.0 & 0.0 & 1.0 & 132.9 & 4.14 & 22.71 & 60.82 \\
 & UGC11910 & 331.76 & 10.23 & 0.0 & 0.0 & 1.0 & 60.8 & 2.34 & 21.63 & 79.23 \\
 & NGC4388 & 186.44 & 12.66 & 0.0 & 0.0 & 1.0 & 323.0 & 2.55 & 22.78 & 78.43 \\
 & Mrk3 & 93.9 & 71.04 & 0.0 & 0.0 & 1.0 & 279.4 & 7.98 & 37.09 & 94.26 \\
 & LEDA168563 & 73.02 & 49.55 & 0.0 & 0.0 & 1.0 & 58.7 & 4.28 & 22.35 & 62.87 \\
 & 3C445 & 335.96 & -2.1 & 0.0 & 0.0 & 1.0 & 55.2 & 1.77 & 22.88 & 96.98 \\
FSRQ & 3C454.3 & 343.49 & 16.15 & 3.71 & 2.01 & 0.150 & 124.9 & 4.54 & 35.05 & 114.93 \\
 & 3C273 & 187.28 & 2.05 & 3.51 & 2.20 & 0.230 & 434.7 & 2.88 & 33.35 & 133.67 \\
BL Lac & Mrk421 & 166.11 & 38.21 & 0.0 & 0.0 & 1.0 & 182.9 & 3.86 & 22.84 & 61.88 \\
\enddata
\tablecomments{The table contains the results from the individual source search. It shows the sources, their type, their positions in 
equatorial coordinates (J2000), the best-fit number of astrophysical neutrino events $\hat{n}_s$, the best-fit
spectral index $\hat{\gamma}$, pre-trials $p$-value, the 90\% CL flux upper limit per flavor for $\gamma = 2.0$, 2.5 and 3.0 at a normalization energy of 1~TeV given by $\phi^{2.0}_{90\%}$, $\phi^{2.5}_{90\%}$ and $\phi^{3.0}_{90\%}$ in units of $10^{-13}$~TeV$^{-1}$~cm$^{-2}$~s$^{-1}$. X-ray flux in the table is the intrinsic hard X-ray flux in the energy range 14~-~195~keV and in units of 10$^{-12}$~erg~cm$^{-2}$~s$^{-1}$. The source type is obtained from the BASS DR-2 catalog and differs from the 105-month BAT survey classification for some sources, e.g., NGC 1275, Mrk 348, and 3C 120 are classified as blazars instead of Seyferts in the 105-month BAT survey.}
\end{deluxetable}
\capstarttrue